\documentclass[11pt]{article}
\usepackage{epsf}
\usepackage{amsmath}
\usepackage{graphicx}
\usepackage{color}
\usepackage{psfrag}
\usepackage{amssymb}
\usepackage{bm,epsfig,amsmath,amssymb,amsfonts,
latexsym,comment,cancel,verbatim,epsfig}
\usepackage[font=md,captionskip=8pt]{subfig}
\setcaptionwidth{14cm}

\usepackage{ifpdf}

\usepackage{graphics}

\newcommand{\cD}{{\cal D}}

\newcommand{\cL}{{\cal L}}

\newcommand{\cO}{{\cal O}}

\newcommand{\Tr}{\mbox{Tr}}

\newcommand{\ra}{\rightarrow}
\newcommand{\be}{\begin{equation}}
\newcommand{\ee}{\end{equation}}
\newcommand{\bea}{\begin{eqnarray}}
\newcommand{\eea}{\end{eqnarray}}

\newcommand{\bmat}{\left(\begin{array}}
\newcommand{\emat}{\end{array}\right)}

\def\yzero{\smash{\hbox{$y\kern-4pt\raise1pt\hbox{${}^\circ$}$}}}

\def\beq{\begin{equation}}
\def\eeq{\end{equation}}
\def\beqa{\begin{eqnarray}}
\def\eeqa{\end{eqnarray}}

\def\-{\hphantom{-}}

\def\s2{\frac{1}{\sqrt2}}

\def\beq{\begin{equation}}
\def\eeq{\end{equation}}
\def\beqa{\begin{eqnarray}}
\def\eeqa{\end{eqnarray}}

\def\Tr{{\rm Tr \,}}

\def\nn{\nonumber}

\def\IF{\relax{\rm I\kern-.18em F}}
\def\II{\relax{\rm I\kern-.18em I}}

\def\Dsl{\,\raise.15ex\hbox{/}\mkern-13.5mu D} 

\def\CO{{\cal O}}

\catcode`\@=11   
\newdimen\@rotdimen
\newbox\@rotbox  

\def\@vspec#1{\special{ps:#1}}
\def\@rotstart#1{\@vspec{gsave currentpoint currentpoint translate
   #1 neg exch neg exch translate}}
\def\@rotfinish{\@vspec{currentpoint grestore moveto}}
\def\@rotr#1{\@rotdimen=\ht#1\advance\@rotdimen by\dp#1%
   \hbox to\@rotdimen{\hskip\ht#1\vbox to\wd#1{\@rotstart{90 rotate}%
   \box#1\vss}\hss}\@rotfinish}
\def\@rotl#1{\@rotdimen=\ht#1\advance\@rotdimen by\dp#1%
   \hbox to\@rotdimen{\vbox to\wd#1{\vskip\wd#1\@rotstart{270 rotate}%
   \box#1\vss}\hss}\@rotfinish}%
\def\@rotu#1{\@rotdimen=\ht#1\advance\@rotdimen by\dp#1%
   \hbox to\wd#1{\hskip\wd#1\vbox to\@rotdimen{\vskip\@rotdimen
   \@rotstart{-1 dup scale}\box#1\vss}\hss}\@rotfinish}%
\def\@rotf#1{\hbox to\wd#1{\hskip\wd#1\@rotstart{-1 1 scale}%
   \box#1\hss}\@rotfinish}%
\def\rotate{\@ifnextchar[{\@rotate}{\@rotate[l]}}
\def\@rotate[#1]#2{\setbox\@rotbox=\hbox{#2}\@nameuse{@rot#1}\@rotbox}

\catcode`\@=12

\usepackage[a4paper,hmarginratio=1:1,vmarginratio=2:3,totalwidth=15.5cm,
totalheight=22cm]{geometry}

\begin{document}

\makeatletter
\@addtoreset{equation}{section}
\makeatother
\renewcommand{\theequation}{\thesection.\arabic{equation}}

\begin{flushright}
\hfill{}\\
\end{flushright}

\pagestyle{plain}
\thispagestyle{empty}
\setcounter{page}{0}
\renewcommand{\thefootnote}{\arabic{footnote}}
\setcounter{footnote}{0}

\vspace{1.5cm}
\begin{center}
{\bf\LARGE Higgs diphoton rate enhancement from 

\bigskip
supersymmetric physics beyond the MSSM}


\vspace{1.5cm}

\parbox{15.9cm}{
{\large{\bf Marcus Berg$^{\,a}$,  Igor Buchberger$^{\,a}$, D.\,M. Ghilencea$^{\,b,c}$,
 Christoffer Petersson$^{\,d,e}$}}}
\vspace{0.2cm}

{
$^{a}$
Department of Physics, 
Karlstad University, 651 88 Karlstad, Sweden.\\[2mm]
$^{b}$ 
Theory Division, CERN, 1211 Geneva 23, Switzerland.\\[2mm]
$^c$
Theoretical Physics Department, National Institute of Physics\\
and Nuclear Engineering (IFIN-HH) Bucharest MG-6, 077125 Romania.\\[2mm]
$^{d}$ 
 Physique Th\'eorique et Math\'ematique, Universit\'e Libre de Bruxelles,\\
 C.P. 231, 1050 Bruxelles, Belgium.\\[2mm]
$^{e}$ 
International Solvay Institutes, Brussels, Belgium. 
}
\end{center}


\vspace{8mm}

\begin{abstract}
\noindent
We show that supersymmetric ``new physics'' beyond the MSSM
can naturally accommodate a Higgs mass near 126 GeV and
 enhance the signal rate in the $h\to \gamma\gamma$ channel, while 
the signal rates in all the other Higgs decay  channels coincide 
 with Standard Model expectations, except possibly the  $h\to Z\gamma$ channel. 
The ``new physics'' that corrects the relevant Higgs couplings
can be captured by two supersymmetric 
effective operators. We provide a simple example of an underlying model in which
these operators   are simultaneously  generated.
 The  scale of ``new  physics'' that generates these operators can be around
5 TeV or  larger, and  outside the reach of the LHC.
\end{abstract}
\ \\ \ \\
\hrule 
\ \\
{\small
 marcus.berg@kau.se, igorbuch@kau.se, dumitru.ghilencea@cern.ch, christoffer.petersson@ulb.ac.be  
}
\newpage

\section{Introduction}\label{Intro}

The ATLAS and CMS collaborations at CERN  recently presented strong experimental evidence 
for a Higgs-like resonance around 126 GeV \cite{:2012gk}, marking 
a historic achievement in particle physics.  
The signal rates in the $ZZ^\ast$ and $WW^\ast$ channels are in good agreement 
with the Standard Model (SM) predictions. The $b\bar{b}$ and $\tau^+\tau^-$
 signal rates are also compatible with the SM, 
although with substantial error bars. 
At the time of writing, the signal rate in the $h \rightarrow \gamma \gamma$ channel 
is  about $1.5-2$ times larger  than the SM prediction. This discrepancy is only
 at the level of two
 standard deviations and there are theoretical uncertainties \cite{Baglio:2012et}. 
Nevertheless, it is
 interesting to contemplate whether  physics beyond the SM (BSM) can be responsible 
for this discrepancy (for recent work in this direction, see \cite{diphoton}).

In this work we shall assume that the excess in the diphoton channel is due 
to BSM physics that has negligible effect on channels other than
 ${h\rightarrow \gamma\gamma}$. Indeed, this channel is
sensitive to new physics, since it is a loop-level process in the SM.  
With this in mind,  we shall focus on 
the minimal supersymmetric (SUSY) extension of the SM (MSSM) close to the ``decoupling limit''\cite{Gunion:2002zf}, in which the lightest neutral CP-even Higgs boson $h$ is SM-like.

It  is surprising that, despite its large number of parameters, the MSSM has difficulties in
accommodating an enhancement of the ${h\rightarrow \gamma\gamma}$ partial decay width $\Gamma_{h\gamma\gamma}$ without affecting the other partial decay widths. In fact this requirement seems to single 
out loop-induced contributions from  very light color singlet superpartners with a significant coupling to the Higgs,  meaning strongly mixed light stau sleptons, at around 100 GeV \cite{stau}.
However, this introduces issues with vacuum stability and may even be possible to rule out 
at the LHC. In addition, large radiative corrections are needed
to obtain a mass of $m_h\approx$ 126 GeV for the lightest Higgs of the MSSM. 
This requires large supersymmetry breaking terms, such as TeV stop masses and/or a 
large top $A$-term. The lack of evidence for superpartners in the direct SUSY searchers at the LHC
also indicates that soft terms should be large.  However, 
 large supersymmetry breaking terms lead to severe  fine-tuning \cite{Ghilencea:2012gz,Hall:2011aa} 
 in most versions of the MSSM\footnote{
For a discussion of the negative impact of the EW 
fine tuning on the $\chi^2$ fit of such models see also  \cite{Ghilencea:2012qk}.}
 (with or without universal gaugino masses, or 
Higgs soft masses  different from $m_0$ and also from each other).
 This situation suggests that a solution to this 
problem  is not in the SUSY breaking sector but rather in the one that preserves 
supersymmetry.
Following this idea, the purpose of this work is to answer whether one can 
have a minimal {\it supersymmetric}  extension of the Higgs sector that 
allows $m_h\approx 126$ GeV  without undue fine tuning 
and a {\it simultaneous} $h \rightarrow \gamma\gamma$  enhancement, while complying 
with the negative SUSY searches so far.

To this end, one way to proceed is suggested by minimal extensions of the MSSM
Higgs sector, like the Next-to-MSSM (NMSSM) model
which contains an additional singlet chiral superfield (see \cite{Ellwanger:2009dp} 
for a review), and by models where the soft $B\mu$ term is promoted to a SUSY operator 
\cite{lowscale}. In the NMSSM an enhancement of the 
$h\rightarrow\gamma\gamma$ branching ratio  is possible (although this also alters other 
couplings) beyond the SM level \cite{king}. However,  the NMSSM 
remains badly fine-tuned (fine tuning $\Delta>200$) for $m_h\approx 126$ GeV \cite{Ross:2011xv}.
There are known ways to bypass this problem such as in the
 so-called ``generalized'' version of NMSSM (GNMSSM) 
with a superpotential mass term for the singlet superfield, where
the  electroweak fine tuning is significantly reduced to more acceptable levels 
($\Delta\approx 30$) 
\cite{Ross:2011xv,Cassel:2009ps,Delgado:2010uj}. 
These are examples of explicit supersymmetry-preserving modifications of the MSSM that can 
render  more natural the interpretation of the resonance at 126 GeV as the lightest Higgs. 

In this work, instead of considering such specific models, we shall
relax the rigid, minimal structure of the MSSM Higgs sector
and perform an effective field theory analysis of the most relevant
SUSY-preserving operators in this sector. This approach should recover,
in a particular region of the parameter space, scenarios such as those presented above
(GNMSSM, etc).
 We show that it is possible to naturally accommodate a Higgs mass of 126 GeV, 
an enhanced Higgs coupling to photons, and {\it simultaneously}  SM-like Higgs
 couplings to the other particles, 
using only a few supersymmetry-preserving operators  with small coefficients.

In general there is a large set of operators that one could consider in 
the Higgs sector \cite{Piriz:1997id}. Regarding the Higgs mass it is known that
 the presence of the effective dimension 5 superpotential operator \cite{Brignole:2003cm,Dine}
\bea
\label{mHop}
 \frac{1}{M}\,(H_u\cdot H_d)^2~
\eea

\medskip\noindent
can accommodate $m_h\approx 126$ GeV 
without undue fine tuning  \cite{Cassel:2009ps}. The suppression scale $M$  represents 
the mass scale of the SUSY degrees of freedom that have been integrated out to generate \eqref{mHop}. 
Concerning $\Gamma_{h\gamma\gamma}$, from the list of effective operators of dimensions $d=5$ and $d=6$
in the Higgs sector \cite{Cassel:2009ps,extraops} one notices the presence of a SUSY  effective operator 
\medskip
\bea
\label{Hggop}
 \frac{1}{M^2}\,(H_u \cdot H_d)\, \Tr (W^{\alpha } W_{\alpha })
\eea

\medskip\noindent
that can significantly modify the $h\rightarrow \gamma\gamma$ rate. 
Here $W_{\alpha }$ is the electroweak gauge field strength superfield. 
Based on these observations we intend to investigate closer 
the phenomenological impact of (\ref{mHop}) and (\ref{Hggop}).
As a simple example will show, these operators can  be  
generated simultaneously by an underlying microscopic model, making their combination
rather natural. The effect of both operators is maximized for small $\tan\beta$, 
where the MSSM tree level contribution to the Higgs mass is  minimized. 
This means that the impact of the $d=5$ operator  in accommodating  an $m_h\approx 126$ GeV 
 is rather significant.  To our knowledge, 
 the particular combination of the effective SUSY  operators in \eqref{mHop} and 
\eqref{Hggop} has not been studied in the past for the problems we address.\footnote{The 
operator in \eqref{Hggop} was separately discussed in \cite{Heckman:2012nt}.}

The paper is organized as follows.
In Section \ref{sectionL} we  calculate the 
corrections from the effective operators to the Higgs mass and mixing angle. 
In Section \ref{sectionHcoupl} 
we discuss how these operators correct the Higgs couplings and signal rates, 
with focus on the decoupling 
limit. The results in terms of the Higgs mass and the partial widths for the $h\ra \gamma\gamma$ 
and $h\ra Z\gamma$ channels are discussed in Section \ref{results}. 
Section \ref{sec:micro} provides an example of the origin of the effective operators and  Section \ref{Conclusions} 
contains our conclusions, while some details concerning the on-shell
Lagrangian are given in the Appendix.

\section{Corrections from SUSY operators to the Higgs couplings}\label{sectionL}

The effective model we consider consists of the usual MSSM Higgs sector, extended by the operators discussed in the introduction. The relevant part of the
Lagrangian is, in  standard notation,
\bea
\cL=\int\! d^4\theta
\!\sum_{i=u,d} \left( 1-{m}_i^2\theta^2\bar{\theta}^2 \right)
\,H_i^\dagger e^{V_i} H_i
+\Big(
\!\int\!d^2\theta\, \mu \, (1+B\, \theta^2)\, H_d \cdot H_u
+\mbox{h.c.}\Big)+\mathcal{O}_5+\mathcal{O}_6
\eea
where the chiral superfields have components $H_i\equiv (h_i,\psi_i,F_i)$, and
 ${m}_i$ and $B$ are the soft terms. 
$\mathcal{O}_5$ is the only operator of dimension 5 
that one can write  in the Higgs sector,
 up to non-linear field  redefinitions \cite{Antoniadis:2008es}, and 
 has the form:
\medskip
\begin{eqnarray} 
\mathcal{O}_5& = & \frac{c_0}{M}\int d^{2}\theta \,
\,(H_u\cdot H_d)^{2} + \mbox{h.c.}
\label{dim5}
\end{eqnarray}

\medskip\noindent
 For  the component fields expression of $\mathcal{O}_5$
see eq.\ \eqref{dim5a}.

There is a long list of  operators in the Higgs sector of dimension $d=6$
\cite{Cassel:2009ps,extraops,Antoniadis:2008es}. 
A careful analysis of these operators shows that of all these
there is one of them that can couple, in a supersymmetric way, to two gauge bosons:
\medskip
\bea\label{dim6}
\mathcal{O}_6 &=&\frac{1}{M^{2}}\sum_{s=1,2}\frac{c_s}{16 g^2_s \kappa_s}
\int d^{2}\theta 
\,{\rm Tr}( W^{\alpha } W_{\alpha } )_s
(H_u\cdot H_d)+\mbox{h.c.} 
\eea

\medskip\noindent
Here  $g_1$ and $g_2$ denote the U(1)$_Y$
and SU(2)$_L$ gauge couplings,
respectively, and $\kappa_s$ is  a constant that 
cancels the trace factor. $W^\alpha$ is the SUSY field strength
of the U(1)$_Y$ (SU(2)$_L$) vector superfield $V_1$ ($V_2$) of
components $(\lambda_s, V_{s,\mu}, D_s/2)$, $s=1,2$.
$\mathcal{O}_5$ and $\mathcal{O}_6$ provide a minimal set of operators that 
is enough for our purposes.
One can also consider SUSY breaking effects associated to these operators
(see the appendix, eq.\ \eqref{dim6a}), but we only seek supersymmetric solutions
to our problem. The effective expansion is reliable when  $c_{0,1,2}=\cO(1)$ and
$M$ is the largest scale in the theory. 
One can choose one of
$c_{0,1,2}$, for example $c_0$, and set it to $c_0=1$ by redefining $M$.
But it is useful to keep $c_0$ to easily trace or turn off the effects of $\mathcal{O}_{5}$.
Also, to modify the diphoton rate $c_1$ or $c_2$ (or a combination thereof) is enough, 
so together with the scale $M$ we  effectively have only two parameters.
Keeping both $c_{1,2}$  generates an
additional interesting coupling, see later.

 Additional operators of $d=6$ can be present.
Although they could have an impact on the Higgs mass \cite{extraops}, they  have
an additional scale suppression relative to\footnote{
Strictly speaking, this is true for  small $\tan\beta$ region, that will actually be 
the relevant region in our case.} $\mathcal{O}_5$. 
There is an operator similar to $O_6$  
but involving instead the SU(3)${}_C$ gauge group,  that we do not consider here;
this  would change dramatically
the Higgs decay rate to gluons, away from  the SM values. We take the 
good agreement with the SM in most channels as evidence that if present, 
the coefficient of this operator must be small. Finally, another reason
to restrict our analysis to $\mathcal{O}_{5,6}$ is that, as 
discussed later, they can 
be simultaneously generated by underlying physics.

\subsection{The on-shell Lagrangian}

The calculation of the on-shell Higgs Lagrangian
extended by $\mathcal{O}_{5}$ and ${\mathcal O}_6$ is detailed in the appendix.
The result is
\bea
\cL\!\!\!&=&
-\,\frac{1}{2}\,\,\left[
 D_2^a D_2^a\,\left(1+ \frac{c_2}{2M^2} \left(h_u\cdot h_d+
 \mbox{h.c.}\right)\right)
+(2\rightarrow 1)\right]
\nonumber\\
&-&
\left| \mu+2\,\frac{c_0}{M}\,h_d\cdot h_u\right|^2\,\,
\big(\vert h_d\vert^2+\vert h_u\vert^2\big)
+\big[
\frac{\mu}{4}\,\left(\frac{c_2}{M^2}\,\lambda^a_2\,\lambda^a_2
+\frac{c_1}{M^2}\lambda_1^2\right)
\big(\vert h_d\vert^2+\vert h_u\vert^2\big)
+\mbox{h.c.}\big]
\nonumber\\
&+&
\Big\{
\frac{c_2}{4 M^2}\, (h_u\cdot h_d)
\big[ i\,(\lambda^a_2 \sigma^\mu{\mathcal D}_\mu\overline\lambda^a_2
  -{\mathcal D}_\mu\overline\lambda^a_2 \overline\sigma^\mu\lambda^a_2)
\big]+\mbox{h.c.}+(2\rightarrow 1)\Big\}
\nonumber\\
&+&\!\!
\frac{c_0}{M} \big[
2\,(h_u\cdot  h_d)(\psi_d\cdot \psi_u)-(h_u\cdot  \psi_d+\psi_u\cdot h_d)^2
\big]\!+\!{\rm h.c.}
\nonumber\\
&+&\!\!
\Big\{\frac{c_2}{4M^2}\,\,
\Big[-\frac{1}{2}\,(h_u\cdot h_d)\,
(F_2^{a\,\mu\nu}F^a_{2\,\mu\nu}+\frac{i}{2}\,\epsilon^{\mu\nu\rho\sigma}
F^a_{2\,\mu\nu}F^a_{2\,\rho\sigma})
\nonumber\\
&-& \sqrt 2\, (h_u\cdot \psi_d+\psi_u\cdot h_d)\,\sigma^{\mu\nu}\lambda^a_2
F^a_{2\, \mu\nu} - \psi_u\cdot \psi_d\,\lambda^a_2\lambda^a_2\,\,\Big]
+(2\rightarrow 1)
+\mbox{h.c.}\Big\}
\nonumber\\
&+&
 \Big[
\mu\,B\,(h_d\cdot h_u)+\mbox{h.c.}\Big]
-\tilde m_d^2\,\vert h_d\vert^2-\tilde m_u^2\vert h_u\vert^2
\label{Lan}
\eea

\medskip\noindent
where $\tilde m_i^2=m_i^2+\vert \mu\vert^2$, $i=u,d$.
For the explicit form of $D_2^a \,D_2^a$ and $D_1^2$, see eqs.~\eqref{dsq0}, \eqref{dsq}.

Eq.\eqref{Lan}
contains all the information one needs to extract the corrections to the 
Higgs masses and couplings.
In particular, notice the presence of  new, supersymmetric couplings:
\medskip
\bea
-\frac{1}{8} (h_u\cdot h_d)\,\left(\frac{c_2}{M^2} \,\Tr F_2^2+\frac{c_1}{M^2}\,\Tr F_1^2\right)
- \Big| \mu+2\frac{c_0}{M}\,h_d \cdot h_u \Big|^2\, (\vert h_d\vert^2+\vert h_u\vert^2)+\mbox{h.c.}
\eea

\medskip\noindent
which are important below.
There are also direct Higgs-higgsino and higgsino-gaugino couplings 
that can be relevant for dark matter models.
From (\ref{Lan}) we find the Higgs scalar potential $V_h$ 
\bea\label{vvv}
V_h&=&\!\!
\tilde m_d^2 \vert h_d\vert^2+\tilde m_u^2 \vert h_u\vert^2
- \big[\mu\,B \,h_d\cdot h_u + \mbox{h.c.}\big]
+
\frac{g_2^2}{2}\,\vert h_d^\dagger \,h_u\vert^2\,\Big[1+ 
\frac{c_2}{2M^2}(h_d\cdot h_u+\mbox{h.c.})\Big]
\nonumber\\
&+&
\!\!\!\!
\frac{1}{8} (\vert h_d\vert^2-\vert h_u\vert^2)^2\Big[g^2\!+\! \big[(h_d\cdot h_u)\Big(
\frac{g_1^2c_1}{M^2}\! +\!\frac{ g_2^2c_2}{M^2}\Big)\!+\!\mbox{h.c.}\big]\Big]\!
\! +\! 
4 \left| \frac{c_0}{M}\right|^2\,\vert h_d\cdot h_u\vert^2
(\vert h_d\vert^2\!+\!\vert h_u\vert^2)
\nonumber\\[2pt]
&+& \Big[
\Big(2\,\frac{c_0}{M}\,\mu^*\Big)\,(\vert h_d\vert^2\!+\vert h_u\vert^2)\,(h_d\cdot h_u)
 +\mbox{h.c.}\Big],\qquad\qquad\qquad
(g^2\equiv g_1^2+g_2^2) \;,
\eea

\medskip\noindent
which depends on two parameters: $c_0$ from the effective
dimension 5 operator and 
the combination $(g_1^2 c_1+ g_2^2c_2)$ from the effective dimension 6 operator.
Note that last term in the first line above does not contribute to the neutral 
Higgs sector masses.

We also include dominant loop corrections, although they do not play the same 
crucial role they do in the MSSM.  In the  small $\tan\beta$ regime and  for
dominant top Yukawa coupling,  the one-loop and leading two-loop correction to $V_h$ is  
\cite{Higgs1loop},
\begin{eqnarray} 
\label{delta}
\Delta V_h=\frac{g^2}{8}\delta \,\vert h_u\vert^4
\eea
where
\bea
\delta  &\equiv&\frac{3\,h_{t}^{4}}{g^{2}\,\pi ^{2}\,}\bigg[\ln \frac{M_{\tilde{t}
}}{m_{t}}+\frac{X_{t}}{4}+\frac{1}{32\pi ^{2}}\,\Big(3\,h_{t}^{2}-16
\,g_{3}^{2}\Big)\Big(X_{t}+2\ln \frac{M_{\tilde{t}}}{m_{t}}\Big)\ln \frac{M_{
\tilde{t}}}{m_{t}}\bigg]
\nonumber\\[3pt]
X_{t} &\equiv & \frac{2\,(A_{t}-\mu \cot \beta )^{2}}{ M_{\tilde{t}}^{2}}
\,\,\Bigg(1- \frac{(A_{t}-\mu \cot \beta )^{2} }{ 12\,\,M_{\tilde{t}}^{2} }\,\Bigg). 
\end{eqnarray}
with $M_{\tilde{t}}^{2}\equiv m_{\tilde{t}_{1}}m_{\tilde{t}_{2}}$, 
and $g_{3}$ is the QCD coupling.

\subsection{The Higgs mass and mixing angle}

The scalars receive mass corrections from the
 usual one-loop radiative corrections  but now also from the effective operators. 
Here we take the parameters $c_0, c_1, c_2$ to be real.
We find the following result for the mass of the lightest Higgs scalar $h$:
\bea
\label{mh}
m_h^2&=&\frac{1}{2}\,\Big\{
m_A^2+m_Z^2+\delta\,m_Z^2\,\sin^2\beta
- \sqrt w\Big\}+\Delta m_h^2
\eea
where
\bea
w&\equiv &
[ \,(m_A^2-m_Z^2)\,\cos 2\beta+\delta\,m_Z^2\,\sin^2\beta]^2
+\sin^2 2\beta \,(m_A^2+m_Z^2)^2 \; . 
\eea
and where $\Delta m_h^2$ is the contribution 
due to the higher-dimensional operators:
\bea
\Delta m_h^2= 
\left(2\mu\,\frac{c_0}{M}\right)\,s_1 
+ 
\left(2\mu \frac{c_0}{M}\right)^2\,s_2 
+  
\left(\frac{g_1^2c_1}{M^2}+ \frac{g_2^2c_2}{M^2}\right)\,s_3
+\cO\!\left(\frac{1}{ M^3}\!\right) \,
\eea
with
\bea
s_1& =& \,v^2\,\sin 2\beta\,\Big\{1+ \frac{(m_A^2+m_Z^2)}{\sqrt w}\Big\}
\\
s_2&=&\!\!
 \frac{v^4}{4\,\mu^2}\,\sin^2 2\beta+
\frac{v^4}{\sqrt w}\, \Big\{-1+\frac{1}{2\mu^2} (m_A^2+m_Z^2) \sin^2 2\beta\Big\}
+
\frac{1}{w^{3/2}}\,(m_A^2+m_Z^2)^2\,v^4\,\sin^2  2\beta
\nonumber\\[4pt]
s_3&=&\!\!\!
\frac{v^4}{32}\,\sin 2\beta\,+\frac{v^4 \sin 2\beta}{128\,\sqrt w}
\big[ 8 m_A^2-(4+3\delta) m_Z^2+6 \delta m_Z^2 \cos 2\beta
+3 (4 m_A^2-\delta m_Z^2)\cos 4\beta\big]\, \nonumber
\eea

\medskip\noindent
where we kept (small) effects from the interplay  between the effective operators
and the one-loop correction to $V_h$.
The mass of the  CP-odd Higgs boson is
\medskip
\bea
m^2_{A}=\frac{2\,B\,\mu}{\sin 2\beta}\,
-\frac{2\,v^2}{\sin2\beta} \left( \frac{c_0}{M}\,\mu\right)
-\frac{v^4}{32}  \frac{\cos^2 2\beta}{\sin 2\beta}\,   
\left(\frac{g_1^2c_1}{M^2}+ \frac{g_2^2c_2}{M^2}\right)
+\cO\!\left(\!\frac{1}{M^3}\!\right)
\eea

\medskip\noindent
which does not receive one-loop corrections.
The mixing angle  $\alpha$ is given by
\bea
\label{mixangle}
\tan 2\alpha\!\!\!\!
&=&\!\!\!
-\frac{1}{\cD}\,
\Big[(m_A^2\!+\! m_Z^2)\tan 2\beta
\!-
\frac{2 \, v^2}{\cos 2\beta}\left(2\mu\frac{c_0}{M}\right)
\!- \Big(2\frac{c_0}{M}\Big)^2 v^4 \tan 2\beta \\
&& 
\hspace{1cm}+
\Big(\frac{g_1^2c_1}{M^2}+ \frac{g_2^2c_2}{M^2}\Big)
\frac{v^4\,( m_Z^2\,f_Z - m_A^2\,f_A)}{32\,\cD \,\cos^2 2\beta}\Big]
\nonumber 
\eea
with 
\bea
\cD&=&m_A^2-m_Z^2+(\sec 2\beta-1) \,\delta \,m_Z^2/2
\nonumber\\[2pt]
f_Z&=&4\,\cos 2\beta-(2-5 \cos 2\beta+6\cos4\beta-3\cos6\beta)\,\delta/4
\nonumber\\[2pt]
f_A&=&\cos 2\beta+3\,\cos 6\beta
\eea

\medskip\noindent
One can see the corrections to $\tan 2\alpha$ due to the effective operators, that
are used below.

We also note that the new operators
correct the  gauge field kinetic terms when the Higgs fields
receive vevs. The corrected gauge couplings
are the ones that appear in the following.

\section{Corrections to the partial widths of $h\to \gamma\gamma$ and $h\to Z\gamma$}
\label{sectionHcoupl}

In this section we study how the new operators correct the Higgs couplings to the SM particles. 
To this end, we parametrize these corrections in terms of the usual MSSM Higgs couplings.

\subsection{Higgs couplings and signal rates}

The renormalizable part of the Lagrangian for the lightest neutral CP-even 
Higgs scalar $h$ can be written \cite{Hcouplings} as 
\begin{eqnarray}
\label{Ltree}
\mathcal{L}_{\mathrm{ren}}&=&  -c_t 
\frac{m_t}{v}h \,t \,\bar{t}-c_c \frac{m_c}{v}h \,c \,\bar{c}-c_b \frac{m_b}{v}h \,b \,\bar{b}-c_\tau \frac{m_\tau}{v}h \,\tau^+ \,\tau^- \nn \\
&&+  c_Z \frac{m_Z^2}{v}h \,Z^{\mu} \,Z_\mu +  c_W \frac{2m_W^2}{v}h \,W^{+\mu} \,W^{-}_\mu
  \end{eqnarray}
where the dimensionless coefficients are given by,
 \begin{eqnarray}\label{ctree}
c_t = c_c=\frac{\cos\alpha}{\sin\beta}~,~
 c_b=c_{\tau}= -\frac{\sin\alpha}{\cos\beta}~,~
 c_Z=c_W=\sin(\beta -\alpha) \,.
\end{eqnarray}
where the mixing angle $\alpha$ is given in \eqref{mixangle}. In the scenario under consideration all loop corrections to the tree level coefficients in \eqref{ctree} are negligible. The usual SM values for the couplings in \eqref{Ltree} and \eqref{ctree} are obtained in the decoupling limit, in which $\alpha\to\beta-\pi/2$, implying that $\cos\alpha\to \sin\beta$, $\sin\alpha\to -\cos\beta$ and hence, $c_i\to c_i^{\mathrm{SM}} =1$, where $i=t,c,b,\tau,Z,W$.

We work in the limit when   
loop contributions from superpartners and other Higgs scalars are negligible. 
The dimension-five part of the Higgs Lagrangian, which takes into account
 1-loop contributions from SM particles as well as the contributions from 
the effective operators in (\ref{Lan}) can be written as 
\begin{eqnarray}
\label{dim5coup}
\mathcal{L}_{\mathrm{dim5}}  &=& c_g^{\mathrm{loop}}\frac{\alpha_{\mathrm{S}}}{12\pi v}  \,h
\Tr G^{\mu\nu} G_{\mu\nu}+\left(c_\gamma^{\mathrm{loop}} +c_\gamma^{\mathrm{BMSSM}}\right) \frac{\alpha_{\mathrm{EM}}}{8\pi  v}   \, h \,F^{\mu\nu} F_{\mu\nu}\nn \\[5pt]
&&+\left( c_{\gamma Z}^{\mathrm{loop}} +c_{\gamma Z}^{\mathrm{BMSSM}} \right) \frac{\alpha_{\mathrm{EM}}}{4\pi \sin\theta_w \,v}   \, h \,Z^{\mu\nu} F_{\mu\nu} ~.
 \end{eqnarray}
The 1-loop contributions to these coefficients are given by\footnote{  
See also \cite{Zgamma} for additional studies of $h\to Z\gamma$.} \cite{dim5}
\begin{eqnarray}
 \label{coefloop}
 c_g^{\mathrm{loop}} &=&  c_t \mathcal{A}_{g}^{(t)} +c_b \mathcal{A}_{g}^{(b)} 
 \approx 1.03\, c_t-(0.05+0.07i)\,c_b \nn \\
c_\gamma^{\mathrm{loop}} &=&  c_W \mathcal{A}_{\gamma}^{(W)}+ c_t \mathcal{A}_{\gamma}^{(t)} 
\approx -8.36 \,c_W +1.84 \,c_t   \nn \\
c_{Z\gamma}^{\mathrm{loop}} &=&   c_W  \mathcal{A}_{Z\gamma}^{(W)}
+c_t \mathcal{A}_{Z\gamma}^{(t)} 
\approx 5.80 \,c_W - 0.31 \,c_t
 \end{eqnarray}
 where, in the last steps, we have inserted $m_h=126$ GeV in the 1-loop form factors $\mathcal{A}$, whose explicit expressions are given in appendix \ref{form}. In the decoupling limit, where $c_i\to c_i^{\mathrm{SM}} =1$ in \eqref{ctree},
the $c^{\mathrm{loop}}$-coefficients in \eqref{coefloop} approach the values they have in the SM, which, for $m_h=126$ GeV, follow trivially from \eqref{coefloop},
\begin{eqnarray}\label{SMvalues}
c_g^{\mathrm{loop}}\to c_g^{\mathrm{SM}} &\approx&  0.98+0.07i\nn \\
c_\gamma^{\mathrm{loop}}\to c_\gamma^{\mathrm{SM}} &\approx& -6.52 \nn\\
c_{\gamma Z}^{\mathrm{loop}} \to c_{\gamma Z}^{\mathrm{SM}} &\approx& 5.49  
 ~.
 \end{eqnarray}

In order to obtain the $c_\gamma^{\mathrm{BMSSM}}$ and $c_{\gamma Z}^{\mathrm{BMSSM}}$ coefficients in \eqref{dim5}, we extract the following component interactions from the operators in \eqref{dim6},
\begin{eqnarray}
\label{higgs}
\mathcal{O}_6 &\supset& - \sum_{s=1,2} \frac{c_s}{8M^2}\,h_u\cdot h_d
 \left( F^{a\,\mu\nu}_{s} F^{a}_{s\,\mu\nu}  +\frac{i}{2}\epsilon^{\mu\nu\rho\sigma}F^{a}_{s\,\mu\nu}F^{a}_{s\,\rho\sigma} \right) + \mathrm{h.c.} \\
 &\supset& \frac{v\,\cos(\beta+\alpha)}{8M^2} \Big(  
 [c_1\cos^2\theta_w+c_2 \sin^2\theta_w]h F^{\mu\nu} F_{\mu\nu}  
 +2(c_2-c_1)\sin\theta_w\, \cos\theta_w\,h F^{\mu\nu}Z_{\mu\nu}
  \Big) \nn
\end{eqnarray}
where we have used,
\begin{eqnarray}
 h_u\cdot h_d = h_u^+ h_d^- -h_u^0 h_d^0 ~ &,&~ h_i^0 = \frac{1}{\sqrt{2}} (v_i+\mathrm{Re}\,h_i^0+i\mathrm{Im}\,h_i^0) \ , \nn \\
 \mathrm{Re}\, h_d^0 = - \sin\alpha\,h+\cos\alpha\,H ~ &,& ~ \mathrm{Re}\, h_u^0 = \cos\alpha\,h+\sin\alpha\,H \ , \nn \\
 A_{1\,\mu} = \cos\theta_w A_{\mu}-\sin\theta_w Z_{\mu} ~ &,& ~ A^{(3)}_{2\,\mu} = \sin\theta_w A_{\mu}+\cos\theta_w Z_{\mu} \ 
\end{eqnarray} 
and $v_d=v\cos\beta$, $v_u=v\sin\beta$, with $v=$ 246 GeV. Moreover, the hypercharge gauge boson $A_{1\,\mu}$ and the (third component of the) SU(2)${}_L$ gauge boson $A^{(3)}_{2\,\mu}$ have been rewritten in terms of the photon $A_\mu$ and the $Z$ boson $Z_{\mu}$. Note that there is also a dimension 5 operator generated from \eqref{dim6} that involves the Higgs scalar $h$ and two field strengths of the $Z$ boson (as well as an analogous operator involving two field strengths of the $W$ boson). However, since these operators will have couplings comparable to the $\gamma\gamma$ or $Z\gamma$ couplings, but strongly phase space suppressed, we expect them to be irrelevant with respect to the usual dimension 3 Higgs coupling to the $Z$ and $W$ bosons in \eqref{Ltree}. Therefore, we do not consider them.

The contributions to \eqref{dim5coup} from \eqref{higgs} are given by:
 \begin{eqnarray}
 \label{coeffB}
c_\gamma^{\mathrm{BMSSM}} &=& \frac{\pi\,v^2\,\cos(\beta+\alpha)}{M^2\alpha_{\mathrm{EM}}}
 (c_1\cos^2\theta_w+c_2 \sin^2\theta_w) \nn \\
c_{\gamma Z}^{\mathrm{BMSSM}} &=& \frac{\pi\,v^2\,\cos(\beta+\alpha)}{M^2\alpha_{\mathrm{EM}}}
 (c_2-c_1)\sin^2\theta_w \cos\theta_w ~.
 \end{eqnarray}
 In the decoupling limit, where $\cos(\beta+\alpha)\to \sin2\beta$, we see that the coefficients in \eqref{coeffB} are maximized for small $\tan\beta$. 

We can now define the relevant Higgs partial decay widths, normalized 
 to the corresponding SM value, in terms of the dimensionless $c$-coefficients 
in \eqref{ctree}, \eqref{coefloop}, \eqref{SMvalues}, \eqref{coeffB}, 
 \begin{eqnarray}
\frac{\Gamma_{hii}}{\Gamma_{hii}^{\mathrm{SM}}}   = \!
 \left| c_i \right|^2 ,
\frac{\Gamma_{hgg}}{\Gamma_{hgg}^{\mathrm{SM}}}
   =  \left| \frac{c_g^{\mathrm{loop}}}{c_g^{\mathrm{SM}}} \right|^2 ,
\frac{\Gamma_{h\gamma\gamma}}{\Gamma_{h\gamma\gamma}^{\mathrm{SM}}}   =\!  
\left| \frac{c_\gamma^{\mathrm{loop}} +c_\gamma^{\mathrm{BMSSM}}}{c_\gamma^{\mathrm{SM}}} \right|^2 ,
\frac{\Gamma_{h\gamma Z}}{\Gamma_{h\gamma Z}^{\mathrm{SM}}}   =  \!
\left| \frac{c_{\gamma Z}^{\mathrm{loop}} +c_{\gamma Z}^{\mathrm{BMSSM}}}{c_{\gamma Z}^{\mathrm{SM}}} \right|^2
\end{eqnarray}
as well as the corresponding branching ratios (BRs),
 \begin{eqnarray}
\label{defBR}
\frac{\mathrm{BR}_{hii}}{\mathrm{BR}_{hii}^{\mathrm{SM}}}   =  \left| \frac{c_i}{c_{\mathrm{tot}}} \right|^2 &,&
\frac{\mathrm{BR}_{hgg}}{\mathrm{BR}_{hgg}^{\mathrm{SM}}}   =  \left| \frac{c_g^{\mathrm{loop}}}{c_g^{\mathrm{SM}}\,c_{\mathrm{tot}}} \right|^2 \nn \\
\frac{\mathrm{BR}_{h\gamma\gamma}}{\mathrm{BR}_{h\gamma\gamma}^{\mathrm{SM}}}   =  \left| \frac{c_\gamma^{\mathrm{loop}} +c_\gamma^{\mathrm{BMSSM}}}{c_\gamma^{\mathrm{SM}}\,c_{\mathrm{tot}}}\right|^2&,&
\frac{\mathrm{BR}_{h\gamma Z}}{\mathrm{BR}_{h\gamma Z}^{\mathrm{SM}}}   =  \left| \frac{c_{\gamma Z}^{\mathrm{loop}} +c_{\gamma Z}^{\mathrm{BMSSM}}}{c_{\gamma Z}^{\mathrm{SM}}\,c_{\mathrm{tot}}} \right|^2  ~.
\end{eqnarray}
The coefficient $c_{\mathrm{tot}}$ in \eqref{defBR} can be written as,
\begin{equation}
\label{ctot}
\left| c_{\mathrm{tot}} \right|^2  =   \sum_{i=t,c,b,\tau,Z,W}\left| c_i \right|^2 \mathrm{BR}_{hii}^{\mathrm{SM}}+
\left| \frac{c_g^{\mathrm{loop}}}{c_g^{\mathrm{SM}}} \right|^2 \mathrm{BR}_{hgg}^{\mathrm{SM}}
\end{equation}
where we have neglected the contributions from, for example, $h\to\gamma\gamma$ and $h\to Z\gamma$, as well as possible invisible decays. 
 Let us now define the inclusive, as well as the individual gluon-gluon fusion (ggF), vector boson fusion (VBF) and vector boson associated (VH) production cross sections, normalized with respect to the corresponding SM values, 
  \begin{eqnarray}
\label{prod}
\frac{\sigma_{\mathrm{incl}}}{\sigma_{\mathrm{incl}}^{\mathrm{SM}}}&=& \frac{\left| c_g^{\mathrm{loop}}/c_g^{\mathrm{SM}} \right|^2\sigma_{\mathrm{ggF}}^{\mathrm{SM}}+\left| c_V  \right|^2 \left( \sigma_{\mathrm{VBF}}^{\mathrm{SM}}+\sigma_{\mathrm{VH}}^{\mathrm{SM}}\right)}{\sigma_{\mathrm{ggF}}^{\mathrm{SM}}+\sigma_{\mathrm{VBF}}^{\mathrm{SM}}+\sigma_{\mathrm{VH}}^{\mathrm{SM}}},\nn \\
\frac{\sigma_{\mathrm{ggF}}}{\sigma_{\mathrm{ggF}}^{\mathrm{SM}}}&=&\left| c_g^{\mathrm{loop}}/c_g^{\mathrm{SM}} \right|^2~,~
\frac{\sigma_{\mathrm{VBF}}}{\sigma_{\mathrm{VBF}}^{\mathrm{SM}}}=\frac{\sigma_{\mathrm{VH}}}{\sigma_{\mathrm{VH}}^{\mathrm{SM}}}=\left| c_V \right|^2
\end{eqnarray}
 where we have denoted $c_V=c_Z=c_W$, since the Higgs couplings to $Z$ and $W$ bosons coincide in \eqref{ctree}. We can now write, for example, the signal rates in the inclusive and dijet channels of the $h\to\gamma\gamma$ decay mode, again normalized with respect to the SM,   
 \begin{eqnarray}\label{rates}
R^{\mathrm{incl}}_{\gamma\gamma} & = & \frac{\sigma_{\mathrm{tot}}}{\sigma_{\mathrm{tot}}^{\mathrm{SM}}}\frac{\mathrm{BR}_{h\gamma\gamma} }{\mathrm{BR}_{h\gamma\gamma}^{\mathrm{SM}}} \nn \\
R^{\mathrm{dijet}}_{\gamma\gamma}  &=&  \frac{\epsilon^{\gamma}_{\mathrm{ggF}}\left| c_g^{\mathrm{loop}}/c_g^{\mathrm{SM}} \right|^2\sigma_{\mathrm{ggF}}^{\mathrm{SM}}+\epsilon^{\gamma}_{\mathrm{VBF}}\left| c_V  \right|^2  \sigma_{\mathrm{VBF}}^{\mathrm{SM}}+\epsilon^{\gamma}_{\mathrm{VH}}\left| c_V  \right|^2\sigma_{\mathrm{VH}}^{\mathrm{SM}}}{\epsilon^{\gamma}_{\mathrm{ggF}}\sigma_{\mathrm{ggF}}^{\mathrm{SM}}+\epsilon^{\gamma}_{\mathrm{VBF}}\sigma_{\mathrm{VBF}}^{\mathrm{SM}}+\epsilon^{\gamma}_{\mathrm{VH}}\sigma_{\mathrm{VH}}^{\mathrm{SM}}}\,\frac{\mathrm{BR}_{h\gamma\gamma} }{\mathrm{BR}_{h\gamma\gamma}^{\mathrm{SM}}}
\end{eqnarray}
 where the $\epsilon^{\gamma}$-coefficients are the selection efficiencies for the different production modes in the dijet-tag category of final states.

 \subsection{The decoupling limit}
 
 Let us now take the decoupling limit, in which,
 \begin{equation}
\label{ ccccc}
\frac{c_i}{c_i^{\mathrm{SM}}} = 
 \frac{c_g^{\mathrm{loop}}}{c_g^{\mathrm{SM}}}=
 \frac{c_\gamma^{\mathrm{loop}} }{c_\gamma^{\mathrm{SM}}}=
 \frac{c_{\gamma Z}^{\mathrm{loop}} }{c_{\gamma Z}^{\mathrm{SM}}}=1
\end{equation}
where $i=t,c,b,\tau,Z,W$. This implies that  $|c_{\mathrm{tot}}|=1$ in \eqref{ctot} and that,
 \begin{eqnarray}
\frac{\Gamma_{hii}}{\Gamma_{hii}^{\mathrm{SM}}}   =  
\frac{\mathrm{BR}_{hii}}{\mathrm{BR}_{hii}^{\mathrm{SM}}} =
\frac{\Gamma_{hgg}}{\Gamma_{hgg}^{\mathrm{SM}}}   = 
\frac{\mathrm{BR}_{hgg}}{\mathrm{BR}_{hgg}^{\mathrm{SM}}}=1
\end{eqnarray} 
whereas,
 \begin{eqnarray}
 \label{normpdw}
\frac{\Gamma_{h\gamma\gamma}}{\Gamma_{h\gamma\gamma}^{\mathrm{SM}}}  =  
\frac{\mathrm{BR}_{h\gamma\gamma}}{\mathrm{BR}_{h\gamma\gamma}^{\mathrm{SM}}} =
 \left|1+ \frac{c_{\gamma,\mathrm{dec}}^{\mathrm{BMSSM}}}{c_{\gamma}^{\mathrm{SM}}} \right|^2 ~~,~~
\frac{\Gamma_{h\gamma Z}}{\Gamma_{h\gamma Z}^{\mathrm{SM}}}   =  
\frac{\mathrm{BR}_{h\gamma Z}}{\mathrm{BR}_{h\gamma Z}^{\mathrm{SM}}} =
 \left| 1+\frac{c_{\gamma Z,\mathrm{dec}}^{\mathrm{BMSSM}}}{c_{\gamma Z}^{\mathrm{SM}}} \right|^2 ~
\end{eqnarray}
for which the coefficients in \eqref{coeffB} are given by, 
 in the decoupling limit,
 \begin{eqnarray}
 \label{coeffBdec}
c_{\gamma,\mathrm{dec}}^{\mathrm{BMSSM}} &=& \frac{\pi\,v^2\,\sin2\beta}{M^2\alpha_{\mathrm{EM}}} (c_1\cos^2\theta_w+c_2 \sin^2\theta_w) \nn \\
c_{\gamma Z,\mathrm{dec}}^{\mathrm{BMSSM}} &=& \frac{\pi\,v^2\,\sin2\beta}{M^2\alpha_{\mathrm{EM}}} (c_2-c_1)\sin^2\theta_w \cos\theta_w ~.
 \end{eqnarray} 
 In the decoupling limit, the production cross sections in \eqref{prod} are all equal to their SM corresponding SM value,
 \begin{equation}
\label{ dddd}
\frac{\sigma_{\mathrm{incl}}}{\sigma_{\mathrm{incl}}^{\mathrm{SM}}}= 
\frac{\sigma_{\mathrm{ggF}}}{\sigma_{\mathrm{ggF}}^{\mathrm{SM}}}=
\frac{\sigma_{\mathrm{VBF}}}{\sigma_{\mathrm{VBF}}^{\mathrm{SM}}}=
\frac{\sigma_{\mathrm{VH}}}{\sigma_{\mathrm{VH}}^{\mathrm{SM}}}=1~.
\end{equation}
Thus, all signal rates, for any production mode, associated with the channels $h\to ii$, for $i=t,c,b,\tau,Z,W$, as well as $h\to gg$, will be equal to their corresponding SM value. In the $h\to \gamma\gamma$ channel, we see that the signal rates in \eqref{rates} (as well as any other signal rate in the $h\to \gamma\gamma$ channel) will be given by the corresponding normalized partial width,
\begin{equation}
\label{hgam}
R_{\gamma\gamma}=R^{\mathrm{incl}}_{\gamma\gamma}=R^{\mathrm{dijet}}_{\gamma\gamma}=\frac{\Gamma_{h\gamma\gamma}}{\Gamma_{h\gamma\gamma}^{\mathrm{SM}}}  =   \left|1+ \frac{c_{\gamma,\mathrm{dec}}^{\mathrm{BMSSM}}}{c_\gamma^{\mathrm{SM}}} \right|^2 
\end{equation}    
 and the same for the $h\to Z\gamma$ channel,
  \begin{equation}
\label{hZgam}
R_{Z\gamma}=R^{\mathrm{incl}}_{Z\gamma}=R^{\mathrm{dijet}}_{Z\gamma}=\frac{\Gamma_{hZ\gamma}}{\Gamma_{hZ\gamma}^{\mathrm{SM}}}  =   \left|1+ \frac{c_{Z\gamma,\mathrm{dec}}^{\mathrm{BMSSM}}}{c_{Z \gamma}^{\mathrm{SM}}} \right|^2 ~.
\end{equation}    

In summary, in the decoupling limit, all the partial decay widths, except for $\Gamma_{h\gamma\gamma}$ and $\Gamma_{hZ\gamma}$, are all equal to their corresponding SM value. This implies that all the production cross sections, as well as the signal rates in all other channels, are equal to their SM values. Moreover, as seen in \eqref{hgam} and \eqref{hZgam}, the partial decay widths for $h\to \gamma\gamma$ and $h\to Z\gamma$, normalized with respect to the SM values, coincide with the corresponding signal rates. Hence, in this limit, $\Gamma_{h\gamma\gamma}/\Gamma_{h\gamma\gamma}^{\mathrm{SM}}$ and $\Gamma_{hZ\gamma}/\Gamma_{hZ\gamma}^{\mathrm{SM}}$ can be compared directly to the measured signal rates.  
From (\ref{coeffBdec}) notice that if $c_1\!=\!c_2$ one can change
 $\Gamma_{h\gamma\gamma}$ without affecting  $\Gamma_{h Z\gamma}$.

\section{Results}
\label{results}

We can now evaluate the effect of the operators in \eqref{dim5} and \eqref{dim6} on the
 mass of the lightest neutral CP-even Higgs particle $h$ and on the partial decay widths
 $\Gamma_{h\gamma\gamma}$ and $\Gamma_{hZ\gamma}$, that directly correspond to the rates in the 
decoupling limit, as discussed in the previous section. 

\begin{figure}[h]
\begin{center}
\includegraphics[width=0.9\textwidth]{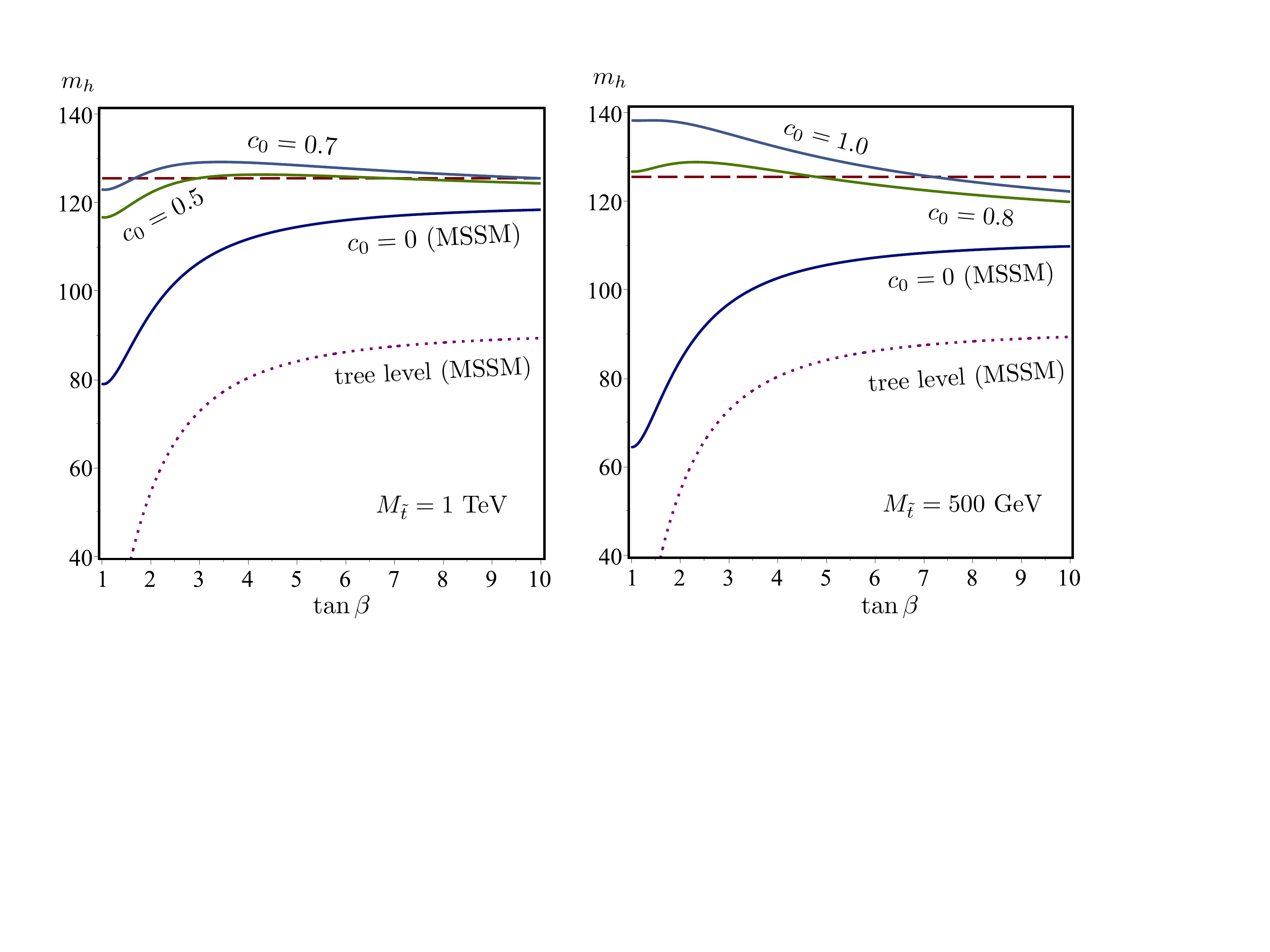}
\end{center}
\vspace{-7mm}
\caption{The mass of the lightest Higgs particle $h$ as a function of $\tan\beta$,
for $M=5 $ TeV, $\mu=300$ GeV, $m_{A}=$ 1 TeV,
no mixing ($X_t=0$), $M_{\tilde t}=$ 500 GeV (left panel) and $M_{\tilde t}=$ 1 TeV (right panel),
and for various values of the ${\mathcal O}_5$ coefficient $c_0$. The solid curves 
include the MSSM loop corrections. }
\label{massplot}
\end{figure}

The Higgs mass in \eqref{mh} as a function of $\tan\beta$ is displayed in figure \ref{massplot}. 
It is well known that we can accommodate a  Higgs mass at 126 GeV
by tuning the soft parameters in the loop correction \eqref{delta},
but this usually demands a large $\tan\beta$, which we do not consider here
(since then additional Yukawa couplings that we do not include become important).
With the $d=5$ operator \eqref{dim5}, one can easily obtain 
a value of $m_h\approx 126$ GeV, see figure \ref{massplot}; this 
has an acceptable fine tuning $\Delta<30$ \cite{Cassel:2009ps},
even for small $\tan\beta<10$, which is an otherwise very fine-tuned 
region of the MSSM. Therefore the effect of the $d=5$ operator
is more important than usually thought.

The dimensionless parameter $\epsilon\equiv c_0\mu/M$ 
measures the extent to which the  contribution from ${\mathcal O}_5$
to the mass can be considered perturbative,
and for the given numbers it is below $\epsilon<0.06$. 
The mass contributions
from ${\mathcal O}_6$ rarely affect noticeably the curves  in figure~\ref{massplot},
 but they are included for completeness (in the figure, $c_1=c_2=-1$).
We included only a subset of the loop corrections, that is relevant at low $\tan\beta$, so
we expect the curves to differ from the complete  result by  few GeV only,
which we confirmed in our examples using {\tt FeynHiggs}~\cite{feynhiggs}.

\begin{figure}[t]
\begin{center}
\includegraphics[width=0.49\textwidth]{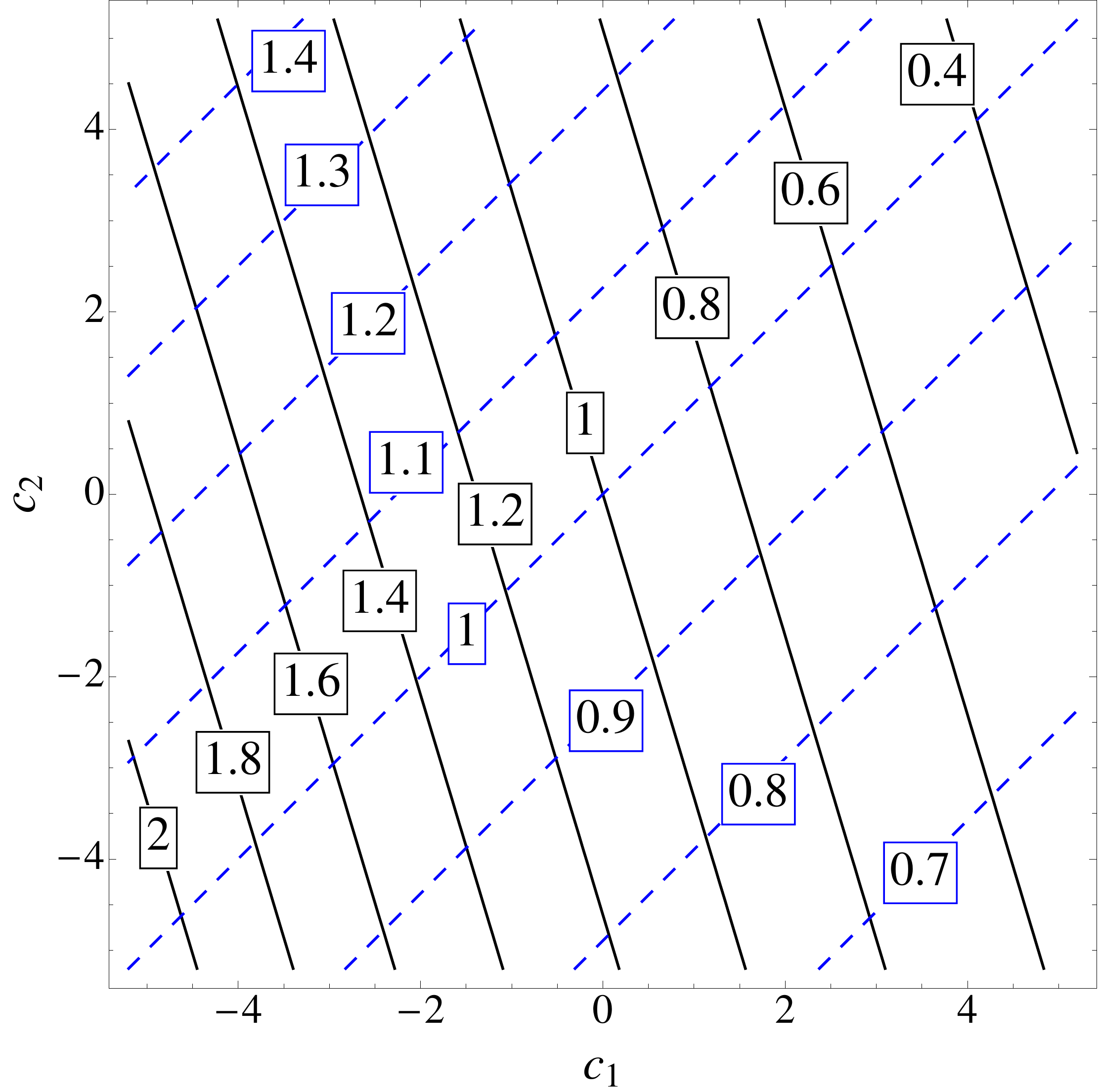}
\includegraphics[width=0.49\textwidth]{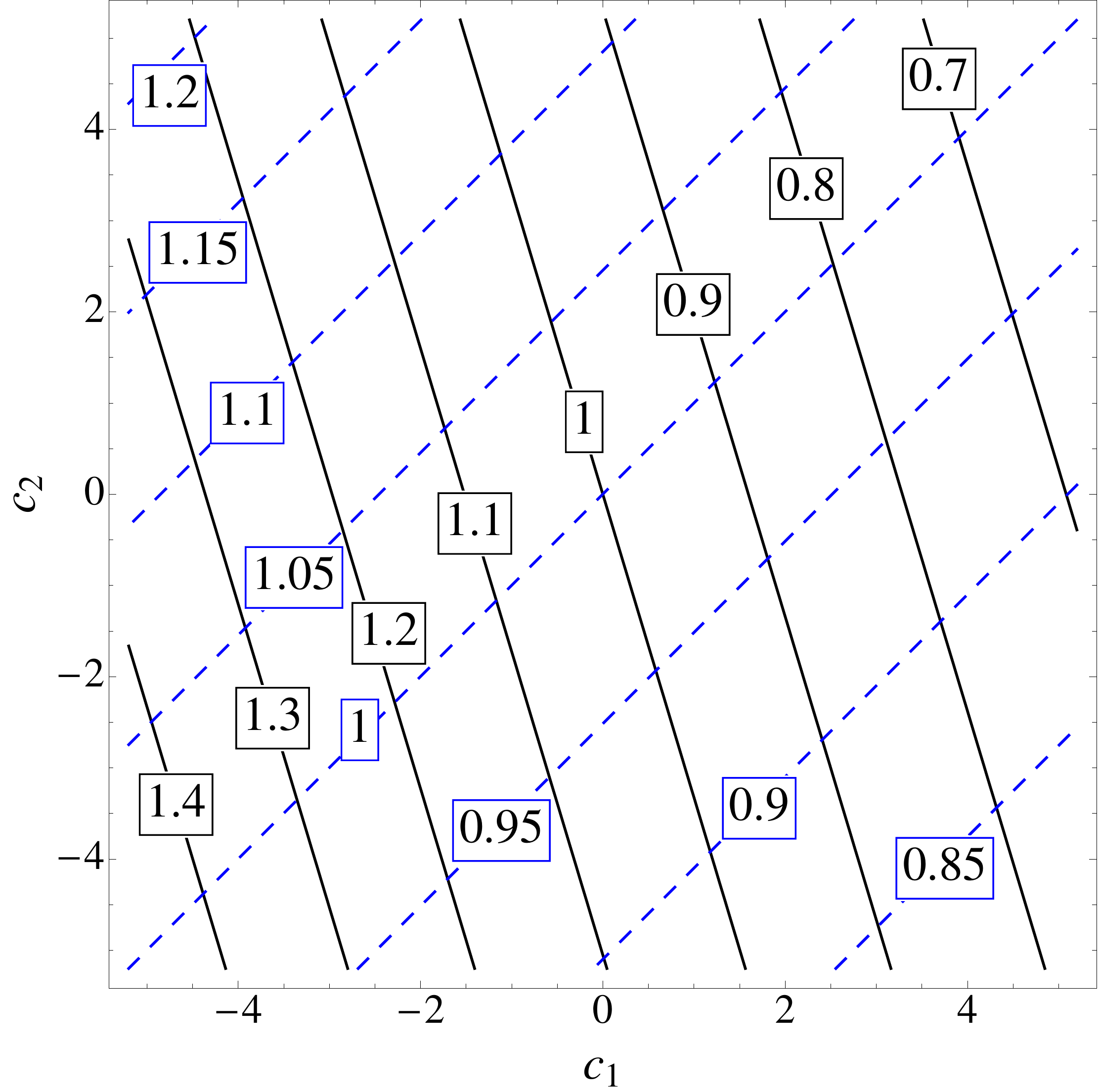}
\end{center}
\caption{The $h\to \gamma\gamma$ signal rate $R_{\gamma\gamma}$ (solid black lines) in 
\eqref{hgam} and the $h\to Z\gamma$ rate $R_{Z\gamma}$ (dashed blue lines) in 
\eqref{hZgam} are shown as functions of the coefficients $c_1$ and $c_2$ of the
 operators in \eqref{dim5} and \eqref{dim6}. In the plots we have set $\tan\beta=3$ 
(left panel) and $\tan\beta=7$ (right panel), $M=5$ TeV and taken the decoupling limit.}
\label{normpdwplot}
\end{figure}

For the value of  the scale $M$ found above in the region of 5 TeV, 
not within the LHC reach,  one would like to examine the signal rates
for $h\to \gamma\gamma$ and $h\to Z\gamma$. These signal rates $R_{\gamma\gamma}$ and $R_{Z\gamma}$
in the decoupling limit, given in eqs.~\eqref{hgam} and \eqref{hZgam}, are shown in figure
 \ref{normpdwplot} as functions of the coefficients $c_1$ and $c_2$ of the operators in
 \eqref{dim5} and \eqref{dim6}. 
Concerning the $h\to \gamma\gamma$ channel, from the dependence on $c_1$ and $c_2$ in
 $c_{\gamma,\mathrm{dec}}^{\mathrm{BMSSM}}$ in \eqref{coeffBdec}, and from the fact that
 $c_\gamma^{\mathrm{SM}}$ is negative in \eqref{SMvalues}, we see that the maximal
 enhancement of $R_{\gamma\gamma}$ in \eqref{hgam} is obtained for  negative values of 
both $c_1$ and $c_2$. In contrast, for positive (and not too large) values of the two
 coefficients, the $h\to \gamma\gamma$ signal is depleted with respect to the SM 
prediction, as can be seen in figure \ref{normpdwplot}. 
We emphasize again that we actually do not  need
both coefficients $c_1$ and $c_2$ (corresponding
to the U(1)${}_Y$ and SU(2)${}_L$ operators in \eqref{dim6}) to achieve enhancement,
we could set e.g.\ $c_2=0$, but the flexibility this additional parameter affords is useful in the next figure.
The maximum allowed enhancement continues to decrease for larger values 
of $\tan \beta$, unless of course if
we simultaneously lower $M$.

From \eqref{coeffBdec} we see that $c_{\gamma Z,\mathrm{dec}}^{\mathrm{BMSSM}}$ is maximized when   $c_1$ and $c_2$ have opposite signs. Moreover, since $c_{\gamma Z}^{\mathrm{SM}}$ is positive in \eqref{SMvalues}, in order to achieve an enhancement of $R_{Z\gamma}$ in \eqref{hgam}, it is required that $(c_2-c_1)>0$. This is seen in figure \ref{normpdwplot}, where $R_{Z\gamma}$ is maximized for large positive values for $c_2$ and large negative values for $c_1$. Notice that the dependence on the sign of $c_1$ and $c_2$ for $R_{\gamma\gamma}$ and $R_{Z\gamma}$ is not specific to this scenario or SUSY. It simply follows from EW symmetry breaking, as can be seen in \eqref{higgs}. 

\begin{figure}[t!]
\begin{center}
\includegraphics[width=1\textwidth]{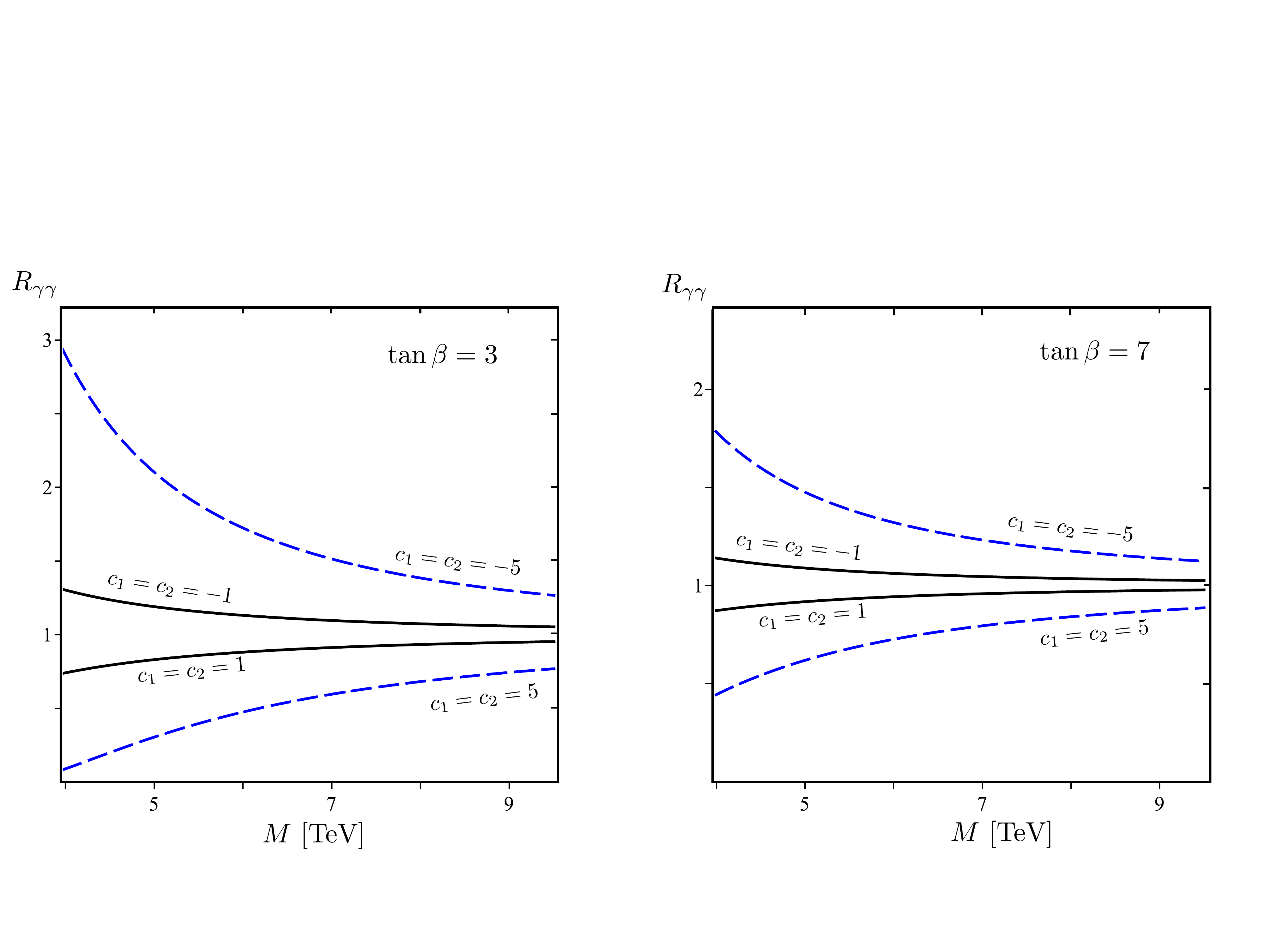}
\end{center}
\caption{The $h\to \gamma\gamma$ signal rate $R_{\gamma\gamma}$ from \eqref{hgam} 
as functions of the scale $M$, where we have set the coefficients of the two operators in \eqref{dim6}  {\it equal},
$c_1=c_2$. We have set  $M_{\tilde{t}}=1$ TeV, $\mu=300$ GeV, $m_{A}=1$ TeV, $X_t=0$.
Dashed blue lines provide rough estimates of the range of validity of the effective field theory.}
\label{ggMplot}
\end{figure}
\begin{figure}[ht!]
\begin{center}
\includegraphics[width=0.9\textwidth]{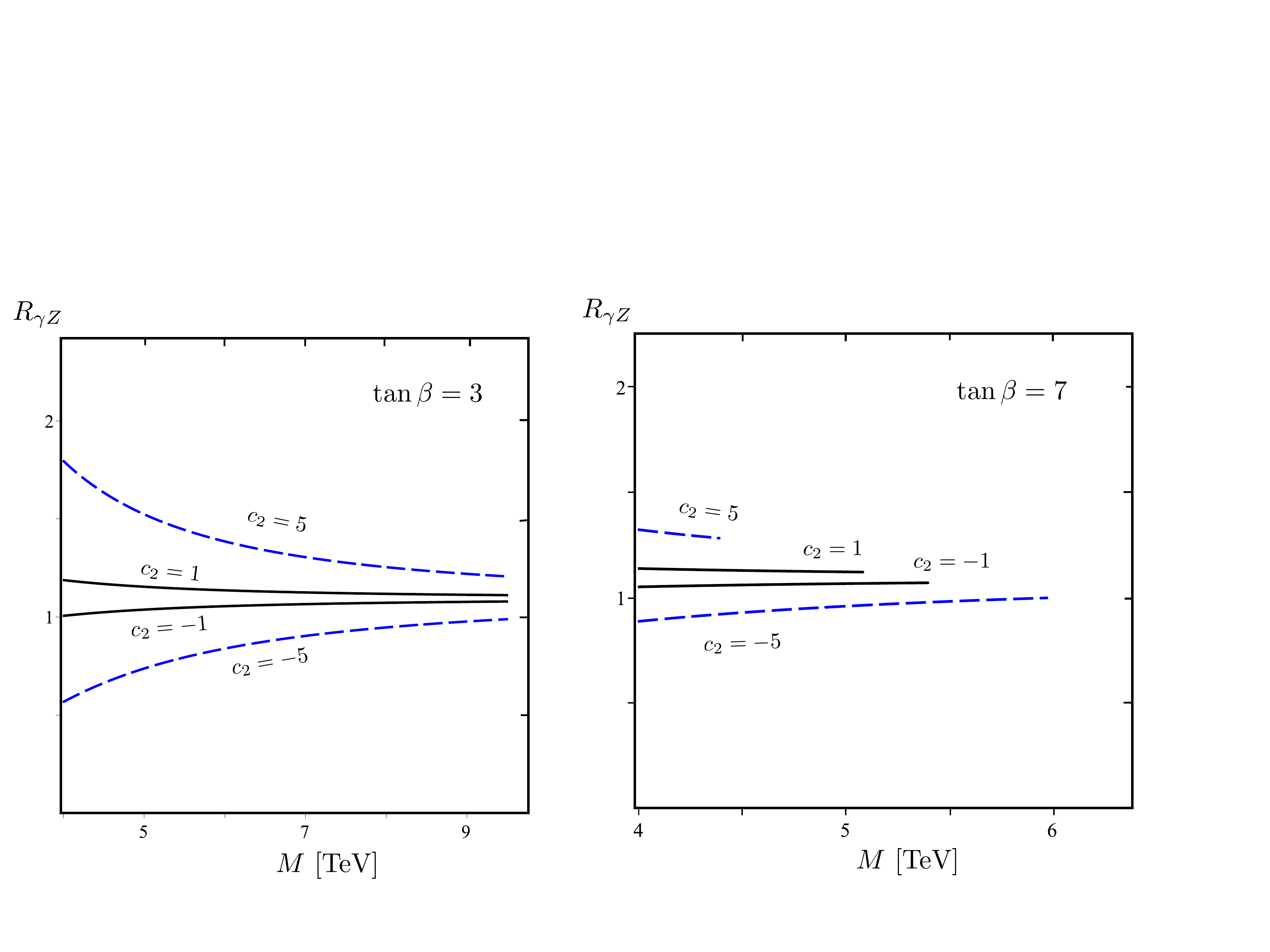}
\end{center}
\caption{The $h\to Z\gamma$ rate $R_{Z\gamma}$ from \eqref{hZgam} for various values of the coefficients $c_1$ and $c_2$ of the operators in \eqref{dim5} and \eqref{dim6}, if we {\it require} $R_{\gamma\gamma}=1.4$
and vary $-5<c_1<5$ for fixed $c_2$.
The curve ends when $c_1$ goes out of range for
$R_{\gamma\gamma}=1.4$ with the given parameters. 
We have set $M_{\tilde{t}}=1$ TeV, $\mu=300$ GeV, $m_{A}=1$ TeV, $X_t=0$.
Dashed blue lines provide rough estimates of the range of validity of the effective field 
theory expansion.}
\label{gZMplot}
\end{figure}

In figures \ref{ggMplot} and \ref{gZMplot}, we show 
a different representation of the
same physics as in figure \ref{normpdwplot}, where we  fix the coefficients $c_1$ and $c_2$ in each curve,
and instead vary the overall scale of new physics $M$. 
As expected,
the effect of the higher-dimensional operators decreases
with increasing $M$, but even for $M$ approaching 10 TeV there
can be some small effect. This perhaps somewhat counter-intuitive behavior
is simply because the relevant SM couplings are small to begin with, 
as emphasized in the introduction. Since the ``new physics'' that
generated these operators comes from a scale around or larger than 5 TeV, 
it will not be within easy reach of the LHC.

In figure \ref{gZMplot} we have  illustrated
the behavior of $R_{\gamma Z}$  if we {\it require} for example $R_{\gamma\gamma}=1.4$
(i.e.\ interpret the diphoton excess as signal)
and vary $-5<c_1<5$, so each curve represents a particular value of $c_2$,
and ends at some upper bound value of $M$ where it is no longer possible to achieve the prescribed
value of $R_{\gamma\gamma}=1.4$. We see that above $\tan \beta=5$ 
or so,   
one  would have  to rely on the scale  $M$ being not too far above 5 TeV.

With hindsight, it may appear that the analysis of the effects of $\CO_5$ in 
\eqref{dim5} and $\CO_6$ in \eqref{dim6} could have been performed mostly independently of each other.
To be clear, we did not assume this: as a matter of principle, we always include the contribution
to e.g.\ the mixing angle $\alpha$ from \eqref{dim5} when computing the effects of \eqref{dim6}.
But we emphasize that the $R_{\gamma\gamma}$ contributions arising from the dimension-six 
operators in \eqref{dim6} are maximized for small $\tan\beta$, see \eqref{coeffBdec}. Therefore, 
since the usual MSSM tree level contribution to the Higgs mass is minimized for small $\tan\beta$, 
the contribution from the dimension-five operator to the tree level Higgs mass is crucial in order to 
accommodate a 126 GeV Higgs mass, as is seen 
\mbox{in figure \ref{massplot}. }

\section{Generating the effective operators from  underlying  physics}
\label{sec:micro}

The natural question is then what ``new physics'' could generate the effective operators 
discussed.
In this section we  discuss a simple example of an underlying model from which 
both $\mathcal{O}_5$ and $\mathcal{O}_2$ arise simultaneously in the low energy effective theory, upon
 integrating out some massive supersymmetric degrees of freedom. Consider a model that contains
 the following superpotential, involving a massive gauge singlet chiral superfield\footnote{Gauge 
singlet fields with a supersymmetric mass term appear in general 
versions of the NMSSM \cite{Cassel:2009ps,Delgado:2010uj,Ross:2011xv}.} $\Sigma$: 
\medskip
\beq
W \, \supset \, 
(\mu+\lambda\, \Sigma)\, H_d\cdot H_u
+\frac{1}{2}\,\mu_S\, \Sigma^2
\eeq
and a gauge kinetic function $\tau$ that depends on $\Sigma$,
\beq
\label{gaugekin}
 \tau(\Sigma)\, \Tr (W_\alpha W^\alpha)\qquad \mathrm{with} \qquad \tau(\Sigma)\supset  
 \frac{\Sigma}{\Lambda}
\eeq
where $\Lambda$ is a dimensionful suppression scale.
If the SUSY mass $\mu_S$ is sufficiently large with respect to the energy scale under 
consideration, then $\Sigma$ can be integrated out supersymmetrically via its 
holomorphic equation of motion, which sets
\beq
 \Sigma=-\frac{\lambda}{\mu_S} H_d\cdot H_u - \frac{1}{\mu_S \Lambda}\Tr (W_\alpha W^\alpha)+\cdots
\eeq
where the dots stand for higher dimensional terms (further suppressed by $\mu_S$, $\Lambda$).
By inserting this solution back into the original Lagrangian, we obtain 
the following terms,
\begin{eqnarray}
&& \int d^2 \theta \left( \mu H_d\cdot H_u - \frac{\lambda^2}{2\mu_S} (H_d\cdot H_u)^2
 - \frac{\lambda}{\mu_S \Lambda} H_d\cdot H_u\,\Tr (W_\alpha W^\alpha) \right)\nn \\
&& + \int d^4 \theta \left( \left| \frac{\lambda}{\mu_S} \right|^2 (H_d\cdot H_u)^\dagger (H_d\cdot H_u) \right)
\label{dim5+6}
\end{eqnarray}
where we have included operators up to dimension-six. 
We see that operators ${\mathcal O}_5$, ${\mathcal O}_6$ 
of \eqref{dim5}, \eqref{dim6} were simultaneously generated as a consequence
 of integrating out $\Sigma$. 

The dimension-six  K\"ahler potential operator in the second line of \eqref{dim5+6} 
gives corrections to the quartic Higgs scalar potential, and hence
to the tree level Higgs mass. However, in comparison  to our dimension-five operator in 
the first line of \eqref{dim5+6}, this operator is  suppressed by a higher power of $\mu_S$. 
As long as $\mu_S$ is sufficiently large, the corrections from this dimension-six  
operator will be smaller in size with respect to the corrections from the dimension-five operator. 

It should be acknowledged that
this example is not renormalizable since the gauge kinetic term \eqref{gaugekin} has 
dimension $d=5$. To have a renormalizable microscopic model, one should also 
specify the degrees of freedom responsible for  generating this $d=5$ operator.
Nevertheless this operator with a moduli-dependent gauge kinetic function
is generically present in models derived
from supergravity or string theory. 

Finally in order to connect with the discussion in the rest of the paper, 
we should assume that the dimension-five operators  in \eqref{gaugekin} only 
involve the U(1)${}_Y$ and SU(2)${}_L$ gauge field strength, and not also  
the SU(3)${}_C$ one that could in principle be also present.
In string models, something along these lines could be achieved by for example 
considering a brane model in which the dimensionality of the branes that give 
rise to the U(1)${}_Y$ and SU(2)${}_L$ gauge
 fields is different from those that give rise to the SU(3)${}_C$ gauge field. In 
this way, since the U(1)${}_Y$ and SU(2)${}_L$ 
branes and the SU(3)${}_C$ branes, respectively, wrap different cycles of the internal 
geometry, they depend on  different gauge singlet moduli fields
 associated with the different cycles. 

\section{Conclusions}
\label{Conclusions}

Recent LHC data on the Higgs mass and its couplings to the photon,
and the negative SUSY searches, present increasing difficulties
 for MSSM-like models 
to naturally accommodate a Higgs mass near 126 GeV, without undue fine-tuning,
and  a potential enhancement of the $h\to \gamma\gamma$ partial decay rate, 
without affecting other partial decay widths. 
Motivated by these observations, in this work we investigated whether
 {\it supersymmetric} effects beyond the MSSM 
 (and even beyond reach for the LHC), could simultaneously accommodate these results. 
Using an effective approach, we identified two effective operators of dimensions $d=5$ and $d=6$
that can address these problems  and  give the leading order corrections to the  Higgs 
quartic coupling and the Higgs coupling to photons, respectively.

We showed that the MSSM  with small, supersymmetric corrections  due 
to these effective operators can 
{\it simultaneously} naturally accommodate a Higgs boson with a mass near 126 GeV, 
an enhanced Higgs coupling to photons (and also $Z\gamma$) relative  to the SM expectation,
 and finally with SM-like Higgs couplings to the other SM particles. 
The scale of the supersymmetric effective operators  is in the region of 5 TeV or
even larger, and therefore possibly not within the LHC reach. The corrections from 
the dimension-six operators to the Higgs coupling to photons (and $Z\gamma$) are 
maximized for  small $\tan\beta$ which is also the region  where the
 dimension-five operator produces the most dramatic
 effect relative  to the MSSM. This suggests that it is natural 
to consider these operators together and this is further  supported by the fact that
both of them can be generated simultaneously by an underlying model, as we showed.

There remains the question of how the existence of these operators can be tested.
Let us assume that  the signal rate in the $h\to \gamma\gamma$ channel is confirmed 
to be higher than the SM expectation while the signal rates in all the other channels 
coincide with the SM values.  If at the same time, 
one can rule out light stau sleptons in the mass range (of 150 GeV or so), 
needed in order to enhance   the diphoton signal with the correct amount,  
this would cause a real problem for the MSSM
and physics beyond the MSSM will be required. 
Should the excess go away when  further data is analyzed, our results will remain useful
to provide bounds on the scale $M$ of  supersymmetric ``new physics'' beyond the MSSM.
Either way, this suggests that  the diphoton rate is a useful, sensitive
 probe in this context.

\bigskip
\section*{Acknowledgements}

The work of C.~Petersson is supported in part by IISN-Belgium (conventions 
4.4511.06, 4.4505.86 and 4.4514.08), by the ``Communaut\'e Fran\c{c}aise de Belgique"
through the ARC program and by a ``Mandat d'Impulsion Scientifique" of the F.R.S.-FNRS.
The work of D.~M.~Ghilencea was supported by a grant of the
Romanian National Authority for Scientific Research, CNCS - UEFISCDI, project number
PN-II-ID-PCE-2011-3-0607.

\appendix
\section{Appendix}
\subsection{Details concerning the Lagrangian}
\label{appendixA}

Here we  derive the Lagrangian of Section~\ref{sectionL}.
Unlike in the text, we also include SUSY breaking effects associated
with operators ${\mathcal O}_5$, $\mathcal{O}_6$, by using the  spurion field. The starting point is
\medskip\bea
\cL=\!\int\! d^4\theta
\!\sum_{i=u,d} \left( 1- m_i^2\theta^2\bar{\theta}^2 \right)
\,H_i^\dagger e^{V_i} H_i+\Big\{
\!\int\!d^2\theta\, \mu \, (1+B \theta^2)\,\, H_d \cdot H_u
\!+\mbox{h.c.}\Big\}+\!\mathcal{O}_5+\mathcal{O}_6\,\,\,
\eea
The  superfield components are $V_i\!=\!(\lambda_i,V_{i,\mu}, D_i^a/2)$, $H_i\!=\!(h_i, \psi_i, F_i)$.
Also $H_d\!\cdot\! H_u=\epsilon^{ij}\,H_d^i\,H_u^j$, with
 $\epsilon^{ij}\,\epsilon^{kj}=\delta^{ik}$;\,\,
$\epsilon^{ij} \epsilon^{kl}=
\delta^{ik} \delta^{jl}-\delta^{il}\delta^{jk}$, $\epsilon^{12}=1$,
 $h_d\!\cdot\! h_u\!=\!h_d^0 h_u^0-h_d^- h_u^+$.
Further 
\begin{eqnarray} 
\mathcal{O}_5\!\!\!\!& = &\!\!
 \frac{1}{M}\int d^{2}\theta \,
\,(c_0 + c_0' \theta^2)\,(H_u\cdot H_d)^{2}\!+\mbox{h.c.} \\
&=& 
\frac{c_0}{M} \,\left[2\, (h_u\!\cdot\! h_d)(h_u\cdot F_d+F_u\cdot h_d-\!\psi_u\cdot \psi_d)
\!-\!(h_u \cdot \psi_d+\!\psi_u \cdot h_d)^2\right]
\!+\!\frac{c_0'}{M}\,(h_u\cdot h_d)^2\!+\!\mbox{h.c.}
\nonumber
\label{dim5a}
\end{eqnarray}
and 
\bea\label{dim6a}
\mathcal{O}_6\!\!\!\!\! &=&\!\!\!
\frac{1}{M^{2}}\sum_{s=1,2}\frac{1}{16 g^2_s \kappa}
\int d^{2}\theta 
\,\,(c_s + c_s' \theta^2) \,{\rm Tr}( W^{\alpha } W_{\alpha } )_s
(H_u\cdot H_d)+ \mbox{h.c.} \\
&=&
\!\!\!\!\!\!\sum_{s=1,2}\!
\frac{c_s}{4M^2} \Big\{\!(h_u\!\cdot\! h_d)
\Big[ i (\lambda^a_s \sigma^\mu{\mathcal D}_\mu\overline\lambda^a_s
\!-\!{\mathcal D}_\mu\overline\lambda^a_s \overline\sigma^\mu\lambda^a_s)
\!+\! D^a_s D^a_s \!-\!\frac{1}{2}
(F_s^{a\,\mu\nu} F^a_{s\,\mu\nu}\!+\!\frac{i\epsilon^{\mu\nu\rho\sigma}}{2} 
F_{s\, \mu\nu}^a F_{s\, \rho\sigma}^a)\Big]\nonumber\\
&-&\sqrt 2\,(h_u\cdot \psi_d+\psi_u\cdot  h_d)
(\lambda^a_s D^a_s+\sigma^{\mu\nu}\lambda_s^a F^a_{s\,\mu\nu})
+(h_u\cdot  F_d+F_u\cdot  h_d-\psi_u\cdot \psi_d)\,\lambda_s^a\lambda_s^a\Big\}
\nonumber\\
&+&
\frac{c_s'}{4M^2} (h_u\cdot h_d)(\lambda^a_s\lambda^a_s)
+\mbox{ h.c.}
\nonumber
\eea

\medskip\noindent
Above we introduced  ${\mathcal D}_\mu\overline\lambda^a=\partial_\mu\overline\lambda^a
-g\,t^{abc}\,V_\mu^b\,\overline\lambda^c$
for covariant derivatives of the gauginos. 

From $\cL$  one finds the equations of motion for the auxiliary fields of Higgs superfields:
\medskip
\bea\label{arhos}
 F_d^{* q}&=&
 -\epsilon^{qp}\,h_u^p\, \left[
 \mu+2\,\frac{c_0 }{ M}\,(h_d\cdot h_u)
-\frac{c_2}{4M^2}\lambda^a_2\lambda^a_2-\frac{c_1 }{4M^2}\lambda_1^2
\right]
 \nonumber\\
 F_u^{* q}&=&
 -\epsilon^{pq}\,h_d^p \,\big[
 \mu+2\,\frac{c_0}{ M}\,(h_d\cdot h_u)
-\frac{c_2}{4M^2}\lambda^a_2\lambda^a_2-\frac{c_1}{4M^2}\lambda_1^2
\big]
 \eea

\medskip\noindent
where $q$ is a SU(2)$_L$ doublet index. 
For the auxiliary fields of the vector superfields we find:
\medskip
 \bea\label{ddd}
 D_2^a&=&-\Big[\, g_2\,(
h_d^\dagger T^a\,h_d\,+h_u^\dagger\,T^a\,h_u)\,
\left(1-\frac{c_2}{2M^2}\,(h_u\cdot h_d+\mbox{h.c.})\right) \label{dd2}
 \\
&&\hspace{2cm}-\frac{\sqrt{2}c_2 }{4M^2}\,\big(\,(h_u \cdot \psi_d+\psi_u\cdot h_d)\,
\lambda^a_2+ \mbox{h.c.}\big)\Big]
\nonumber\\
D_1&=&\!
-\! \Big[ g_1\, (h_d^\dagger \left(\!-\frac{1}{2}\right) h_d +h_u^\dagger
\left(\frac{1}{2}\right) h_u)\, 
\left(1-\frac{c_1}{2M^2}\,(h_u\cdot h_d+\mbox{h.c.})\right)\\
&&\hspace{2cm}-\frac{\sqrt{2}c_1}{4M^2}(h_u\cdot \psi_d+\psi_u\cdot h_d)\,
\lambda_1^a+\mbox{ h.c.}\big)\Big]  \nonumber
\eea
with $T^a=\sigma^a/2$. The squares become
\bea
 D_2^a\,D_2^a&=& \frac{g_2^2}{4}\,
\Big[1-\left( \frac{ c_2}{2M^2}\,h_u\cdot h_d+\mbox{h.c.}\right)\Big]^2
\Big[\,\, (\,\vert h_d\vert^2-\,\vert h_u\vert^2\,\,\big)^2
 +4\,\vert h_d^\dagger\,h_u\vert^2
\Big]
\nonumber\\
&-&
\frac{\sqrt 2}{2} \, g_2 
\Big[\, h_d^\dagger T^a h_d
+h_u^\dagger T^a h_u\,\Big]
\Big[\,
\frac{ c_2}{2M^2}\,(h_u\cdot \psi_d+\psi_u \cdot h_d)\,\lambda^a_2+
\mbox{h.c.}\Big]
\label{dsq0}
\eea
\bea
D_1^2&=&
\frac{g_1^2}{4}\,
\Big[1-\left( \frac{ c_1}{2M^2}\,h_u\cdot h_d+\mbox{h.c.}\right)\Big]^2\,
\big(\,\vert h_d\vert^2- \vert h_u\vert^2\,\big)^2
\nonumber\\&-&
\frac{\sqrt 2}{2} g_1
\Big[ h_d^\dagger \left(\!-\frac{1}{2}\right) h_d +h_u^\dagger
\left(\frac{1}{2}\right) h_u\Big]
\Big[
\frac{c_1}{2M^2}\,(h_u \cdot \psi_d+\psi_u \cdot h_d)\,\lambda_1+
\mbox{h.c.}\Big]
\label{dsq}
\eea

\medskip\noindent
${\mathcal O}_5$ and ${\mathcal O}_6$ and eqs.(\ref{arhos}) to (\ref{dsq})
give the corrections to the MSSM Higgs Lagrangian. Using the corrected
 auxiliary fields in the usual MSSM Higgs Lagrangian, additional terms 
suppressed by $1/M$ and $1/M^2$ are generated.
The full on-shell Lagrangian  is then:
\bea\label{totalL}
\quad 
\cL=\cL_D+\cL_F+\cL_{1}+\cL_{2}
+\cL_{3}+\cL_{\rm SSB} \; .
\qquad \qquad 
\eea
Eliminating the D-dependent terms in $\cL$ one finds,
 see eqs.(\ref{dd2}) to (\ref{dsq}):
\medskip
\bea
\cL_D\!\!\!&=&\!\!\sum_{s=1,2}
-\,\frac{1}{2} \,\, D_s^a D_s^a\,\big[ 
1+ \frac{c_s}{2M^2} (\,h_u\cdot h_d+\mbox{h.c.})\big]
\label{VG}
\eea
and use (\ref{dsq}). Eliminating the $F$-dependent terms in $\cL$ gives $\cL_F$:
\medskip
\bea
-\cL_{F}\!&\equiv&\! \vert F_d\vert^2+
\vert F_u\vert^2
=\vert \mu+2\,\frac{c_0}{M}\,h_d\cdot h_u\vert^2\,\,
\big(\vert h_d\vert^2+\vert h_u\vert^2\big)
\nonumber\\[4pt]
&+&\!\!\!\!
\Big[\mu\,
\left(-\frac{c_2}{4M^2}\lambda^a_2\,\lambda^a_2-\frac{c_1}{4M^2}\,\lambda_1^2\right)
\big(\vert h_d\vert^2+\vert h_u\vert^2\big)
\!+ \mbox{h.c.}\Big]
\label{VF1}
\eea

\medskip\noindent
Apart from auxiliary field contributions, there are also 
terms in the Lagrangian with space-time derivatives, that contribute to the
kinetic terms for Weyl fermions $\psi_{u,d}$, $\lambda^a_{1,2}$ 
when the neutral singlet $h_{u,d}^0$ components of $h_{u,d}$
acquire a vev:
\medskip
\bea
\cL_{1}=
\frac{c_2}{4M^2}\, (h_u\cdot h_d)
\big[ i\,(\lambda^a_2 \sigma^\mu{\mathcal D}_\mu\overline\lambda^a_2
  -{\mathcal D}_\mu\overline\lambda^a_2 \overline\sigma^\mu\lambda^a_2)
\big]+\mbox{h.c.}\, +\left(2\rightarrow 1\right)
\eea

\medskip\noindent
When the Higgs  neutral singlets acquire a vev, these terms produce wavefunction
renormalization of Weyl kinetic terms and a threshold correction to
gauge couplings $g_1$ and  $g_2$.

There are also terms contributing to fermion masses
when the Higgs fields acquire vev's
\bea
\cL_2&=&
\frac{c_2'}{4M^2} (h_u \cdot h_d)(\lambda^a_2\lambda^a_2)
+
\frac{c_1'}{4M^2} (h_u \cdot h_d)(\lambda_1\lambda_1)
\nonumber\\[4pt]
&+&
\frac{c_0}{M}\big[
2\,(h_u\cdot h_d)(-\psi_u\cdot \psi_d)-(h_u \cdot \psi_d+\psi_u \cdot h_d)^2
\big]
+{\rm h.c.}\qquad\qquad\qquad\qquad\qquad
\eea

\medskip\noindent
Further, there are some interaction terms generated
\medskip
\bea
\cL_3\!\!\!
&=&\!\!\!
\Big\{\,\,\frac{c_2}{4M^2}\,\,\Big[- \frac{1}{2}\,(h_u \cdot h_d)\,
(F_2^{a\,\mu\nu}F^a_{2\,\mu\nu}+\frac{i}{2}\,\epsilon^{\mu\nu\rho\sigma}
F^a_{2\,\mu\nu}F^a_{2\,\rho\sigma})
\nonumber\\
&-& \sqrt 2\, (h_u\cdot \psi_d+\psi_u\cdot h_d)\,\sigma^{\mu\nu}\lambda^a_2
F^a_{2\, \mu\nu}
- \psi_u\cdot \psi_d\,\lambda^a_2\lambda^a_2\Big]
+(2\rightarrow  1)
+\mbox{h.c.} \Big\}\qquad
\eea

\medskip\noindent
Finally, the Lagrangian contains ($F$ and $D$ independent)  corrections  
from supersymmetry breaking
due to spurion dependence  in the dimension 5 operator as well as
the usual soft terms of the MSSM. All these together give $\cL_{\rm SSB}$:
\medskip
\bea
\cL_{\rm SSB}=-V_{\rm SSB}\!\!\!&=& \Big[
\frac{c_0'}{M}\,(h_u\cdot h_d)^2
+\mu\,B\,(h_d\cdot h_u)\!+ \mbox{h.c.} \Big]
-\, {m}_d^2 \vert h_d\vert^2-\!{m}_u^2 \vert h_u\vert^2\nonumber
\label{VSSB}
\eea

\medskip\noindent
This concludes the presentation of the Lagrangian to $1/M^2$ order.
From $\cL$ we find the scalar potential $V_h$ for the Higgs sector
shown in the text, eq.~(\ref{vvv}) in which as usual $h_{u,d}$ denote SU(2) doublets.
From this one obtains:
\medskip
\bea
m_h^2&=&m^2_{h,\,\rm loop}+\Delta m_h^2
\eea
with:
\bea
m^2_{h,\,\rm loop}&=&\frac{1}{2}\,\Big\{
m_A^2+m_Z^2+\delta\,m_Z^2\,\sin^2\beta
- \sqrt w\Big\}
\nonumber\\[5pt]
w&\equiv &
[ \,(m_A^2-m_Z^2)\,\cos 2\beta+\delta\,m_Z^2\,\sin^2\beta]^2
+\sin^2 2\beta \,(m_A^2+m_Z^2)^2\,\,\,
\eea
and
\bea
\Delta m_h^2&=&
 f_1\,  (2\mu\frac{c_0}{M})                   
 +f_2\, (-2 \frac{c_0^\prime }{ M})              
+f_3\,  (2\mu\frac{c_0}{ M})^2                       
 +f_4\, (-2 \frac{c_0^\prime}{M})^2             
 +f_5\,  (2\mu\frac{c_0}{ M})(-2 \frac{c_0^\prime}{M})  
\nonumber\\[5pt]
&+&  f_6\, \left(g_1^2\,\frac{c_1}{M^2}+g_2^2 \frac{c_2}{ M^2}\right)
+\cO(\frac{1}{ M^3}) \,
\eea
where
\medskip\bea
f_1& =& v^2\,\sin 2\beta\,\Big\{1+ \frac{(m_A^2+m_Z^2)}{\sqrt w}\Big\}
\nonumber\\
  f_2 &=& \frac{v^2}{2}\,\Big\{1-\frac{\cos 2\beta}{\sqrt w}\,\Big[(m_A^2-m_Z^2)\,\cos 2\beta
  + m_Z^2\,\delta\,\sin^2 \beta\Big]\Big\}
\nonumber\\
f_3&=&
 \frac{v^4}{4\,\mu^2}\,\sin^2 2\beta+
\frac{v^4}{\sqrt w}\, \Big\{-1+\frac{1}{2\mu^2} (m_A^2+m_Z^2) \sin^2 2\beta\Big\}
+
\frac{1}{w^{3/2}}\,(m_A^2+m_Z^2)^2\,v^4\,\sin^2  2\beta
\nonumber\\[4pt]
 f_4&=& -\frac{v^4}{16\,w^{3/2}}\, (m_A^2+m_Z^2)^2\,\sin^2 4\beta
 \nonumber\\
 f_5&=& -\frac{v^4}{4\,w^{3/2}}\, (m_A^2+m_Z^2)\,
 \big[\, \delta \,m_Z^2+(2\,m_A^2-(2+\delta)\,m_Z^2)\,\cos 2\beta\big]\,\sin4\beta
\nonumber\\
f_6&=&\!\!\!
\frac{v^4}{32}\,\sin 2\beta\,\Big[
1\!+\!\frac{1}{4\,\sqrt w}
\,\big[ 8 m_A^2 \!-\!(4\!+\! 3\delta) m_Z^2\!+\! 6 \delta m_Z^2 \cos 2\beta
\!+\! 3 (4 m_A^2\!-\delta m_Z^2)\cos 4\beta\big]\Big]\qquad
\eea

\medskip\noindent
and finally
\medskip
\bea
m^2_{A}=\frac{2\,B\,\mu}{\sin 2\beta}\,
-\frac{\,v^2}{\sin2\beta}\left(2\mu\frac{c_0}{M}\right)
+\left(2\frac{c_0' }{M}\right) v^2    
-\frac{v^4}{32}  \frac{\cos^2 2\beta}{\sin 2\beta}\, \left(\,\frac{g_1^2c_1}{M^2}+ 
\frac{c_2g_2^2}{M^2}\right)
+\cO\left(\frac{1}{M^3}\right) \nonumber
\eea

\bigskip
\subsection{1-loop form factors}\label{form}

The form factors in \eqref{coefloop} are given by,
\medskip
\begin{eqnarray}
\mathcal{A}_{g}^{(t)}=\frac{3}{4} \mathcal{A}_{1/2}(\tau_t) ~ &,&~ 
\mathcal{A}_{g}^{(b)}=\frac{3}{4} \mathcal{A}_{1/2}(\tau_b) \ , \nn \\
\mathcal{A}_{\gamma}^{(W)}=  \mathcal{A}_{1}(\tau_W) ~&,& ~ 
\mathcal{A}_{\gamma}^{(t)}= N_c Q_t^2 \mathcal{A}_{1/2}(\tau_t) \ , \nn \\
 \mathcal{A}_{Z\gamma}^{(W)}= \cos\theta_w\,A_1(\tau_W,\lambda_W) ~&,& ~ 
\mathcal{A}_{Z\gamma}^{(t)}= N_c\frac{ Q_t(2T_3^{(t)}-4Q_t\sin^2\theta_w)}{\cos\theta_w} A_{1/2}(\tau_t,\lambda_t)
\end{eqnarray} 
where $\tau_i=4m_i^2/m_h^2$, $\lambda_i=4m_i^2/m_Z^2$, $N_c=3$, $Q_t=2/3$, $T_3^{(t)}=1/2$ and
\medskip
 \begin{eqnarray}
\mathcal{A}_{1/2}(\tau) & = & 2\tau^{2} \left[\tau^{-1} +(\tau^{-1} -1)f(\tau^{-1})\right]\,  \ ,  \nonumber
\\
\mathcal{A}_1(\tau) & = & -\tau^{2} \left[2\tau^{-2} +3\tau^{-1}+3(2\tau^{-1} -1)f(\tau^{-1})\right]\,  \ , \nn
\\
\mathcal{A}_{1/2}(\tau,\lambda) & = & I_1(\tau,\lambda)-I_2(\tau,\lambda)\ , \nn%
\\%
\mathcal{A}_1(\tau,\lambda) & = & 4(3-\tan^2\theta_w)I_2(\tau,\lambda)+\left[ (1+2\tau^{-1})\tan^2\theta_w-(5+2\tau^{-1}) \right]I_1(\tau,\lambda) 
\end{eqnarray}
where
\bea
I_1(\tau,\lambda)&=&\frac{\tau\lambda}{2(\tau-\lambda)}
+\frac{\tau^2\lambda^2}{2(\tau-\lambda)^2}
\left[f(\tau^{-1})-f(\lambda^{-1})\right]+\frac{\tau^2\lambda}{ (\tau-\lambda)^2}
\left[g(\tau^{-1})-g(\lambda^{-1})\right] \ , \nn
\\%
I_2(\tau,\lambda)&=& -\frac{\tau\lambda}{2(\tau-\lambda)}
\left[f(\tau^{-1})-f(\lambda^{-1})\right] \ ,
\eea
 and
\begin{eqnarray}
f(x)&=&\left\{
\begin{array}{ll}  \displaystyle
\arcsin^2\sqrt{x} & x\leq 1%
\\
\displaystyle -\frac{1}{4}\left[ \log\frac{1+\sqrt{1-x^{-1}}}%
{1-\sqrt{1-x^{-1}}}-i\pi \right]^2 \hspace{0.5cm} & x>1 \ ,
\end{array} \right.  \\[10pt]
g(x)&=&\left\{%
\begin{array}{ll}  \displaystyle
\sqrt{x^{-1}-1}\arcsin\sqrt{x} & x\leq 1
\\%
\displaystyle \frac{\sqrt{1-x^{-1}}}{2}\left[ \log\frac{1+\sqrt{1-x^{-1}}}
{1-\sqrt{1-x^{-1}}}-i\pi \right]^2 \hspace{0.5cm} & x>1 \ .
\end{array} \right.%
\end{eqnarray}

\end{document}